\newcommand{\nombreAutor}{Aurelian Loirette--Pelous}
\newcommand{\tituloTesis}{Dynamical stability of 2D topological lasers}
\newcommand{\directorTesis}{Iacopo Carusotto}
\newcommand{\fechaTesis}{October 2018 - June 2019}
\documentclass[11pt]{report}
\usepackage{setspace}
\usepackage[english]{babel}

\usepackage{lipsum}

\usepackage[letterpaper, top=2.5cm, bottom=2.5cm, left=3cm, right=3cm]{geometry}
\usepackage{fancyhdr}
\fancypagestyle{plain}{
    \fancyfoot[R]{\thepage}

    }

\usepackage{mathptmx}
\usepackage{anyfontsize}

\usepackage[utf8]{inputenc}

\usepackage{graphicx}
\graphicspath{ {images/} } % Specify images path

\usepackage{caption}
\usepackage{amsmath}
\usepackage{footnote}
\usepackage{sidecap}
\usepackage{float}
\usepackage{amssymb}
\usepackage{breqn}
\usepackage[bb=boondox]{mathalfa}
\usepackage{xcolor}
\usepackage{hyperref}
\hypersetup{
    colorlinks,
    citecolor=black,
    filecolor=black,
    linkcolor=black,
    urlcolor=black
}
\usepackage{physics}
\usepackage{multicol}
\usepackage{soul}
\usepackage{esint}

\usepackage[nottoc,notlof,notlot]{tocbibind}
\usepackage{appendix}

\usepackage{titlesec}
\titleformat{\chapter} %section to format
    {\normalfont\fontsize{12}{14.4}\bfseries} %format
    {\thechapter.} %label 
    {12pt} %separation 
    {}
\titlespacing{\chapter}{0pt}{14pt}{0pt}

\titleformat{\section} %section to format
    {\normalfont\fontsize{11}{12.1}\bfseries} %format
    {\thesection} %label 
    {12pt} %separation 
    {}
\titlespacing{\section}{0pt}{14pt}{0pt}

\begin{document}

\begin{titlepage}

\begin{center}
    \includegraphics[width=15cm]{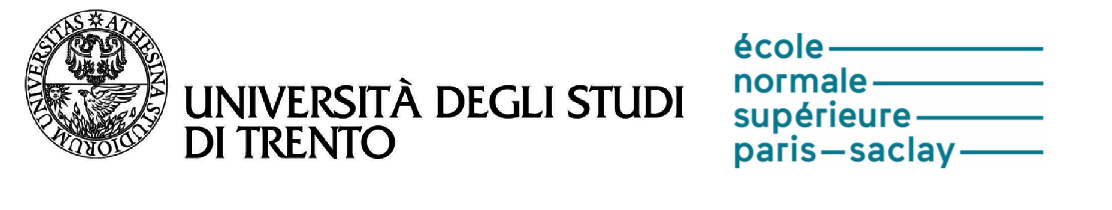} \\
    \vspace{8mm}
    
    {\fontsize{18}{20}
    \selectfont BEC center}   
    
    {\fontsize{14}{16}
    \selectfont \textbf{University of Trento}} \\
    
    \vspace{8mm}
    
    {\fontsize{14}{16}
    \selectfont \textbf{ENS Paris-Saclay}}

    \vspace{8mm}
    
    {\fontsize{14}{16}
    \selectfont "Année de Recherche Pré-doctorale à l'Etranger" report}\\
    
    \vspace{8mm}
    
    {\fontsize{22}{24}
    %\line(1,0){250}\\
    \selectfont \textbf{\tituloTesis}} \\
    %\line(1,0){250}
    
    \vspace{8mm}
    
    %{\fontsize{14}{16}
    %\selectfont Rapport de stage} \\
    
    \vspace{8mm}
    
    {\fontsize{14}{16}
    \selectfont Author \\ \textbf{\nombreAutor}} \\

    \vspace{8mm}
    
    {\fontsize{14}{16} \selectfont Under the supervision of\\ \textbf{\directorTesis} } \\
    
    \vfill
    
    {\fontsize{12}{10} \selectfont Trento, Italia \\ \fechaTesis}
    
\end{center}

\end{titlepage}

\tableofcontents

% Lista de tablas
% Lista de figuras

% \input{chapters/chapter00_abstract.tex}

\chapter*{Introduction}
\addcontentsline{toc}{chapter}{Introduction}

\textbf{Basics of topology:}

In condensed matter, the branch which studies topological phases of matter has been growing rapidly since the discovery of the Integer Quantum Hall Effect by Von Klitzing \cite{qhe} in 1980. In this pioneering experiment, the transverse conductivity of a cold two-dimensional electron gas in a strong perpendicular magnetic field was found to be exactly quantized. Theoretical understanding of this phenomenon was soon brought by Thouless (1982,\cite{thouless}) and Kohmoto (1985, \cite{KOHMOTO}), who related the integer appearing in the Hall conductance to a topological invariant of the system called the \textit{Chern number}. This number is a global quantity which can be attributed to any band of a periodic system (thus gaped), and calculated as $C= \frac{1}{2\pi} \oiint_{BZ} \nabla_{\textbf{k}} \times \mel{u(\textbf{k})}{i\nabla_{\textbf{k}}}{u(\textbf{k})}.d\textbf{s}$ where the $\ket{u(\textbf{k})}$ are the eigenvectors of the states in the given band. Deep mathematical reasons make that with this normalization this number is always a relative integer. Also, this integral is calculated on the whole Brilloin Zone (BZ), which makes its globallity apparent. These two last facts are responsible for the difficulty of passing from a topological phase to another, namely change the value of the Chern number of a band. Indeed, this cannot be achieved by smooth distortions of the system, and rather requires the closure of a neighboring gap. In practice this is very difficult to do, which gives to topological properties a great resistance to spatial disorder as fabrication defects.

This last property is also at the heart of another one of the main feature of topology. Indeed, let us now focus on a given gap of a material, such that the Fermi energy is inside this gap and that the sum of the Chern numbers $ N_1 $ of the bands under the Fermi level be different from 0. We then put a finite size sample of this material in contact with another material with $N2 \neq N_1$. The two materials are in different topological phases, so the gaps need to close somewhere. The consequence of that is the appearance at the interface between the two materials of spatially localized modes, called \textit{edge modes}, whose energy is situated in the middle of the gaps, closing it somehow. 

\begin{figure}[H]   
\begin{minipage}[t]{0.4\textwidth}
In a 2D material in contact with an insulator (for example the air), this closure intervenes by a branch of energy states which links the two bands of the material, as showed in Fig. \ref{fig0.1} (a). As the dispersion of these states has a non-null group velocity, these states will propagate unidirectionnaly along the interface of the materials, as in (b). For this reason, we call them \textit{chiral edge modes}.
\end{minipage}
\begin{minipage}[t]{0.6\textwidth}
    \centering
    \raisebox{\dimexpr-\height+\ht\strutbox\relax}{\includegraphics[scale=0.65]{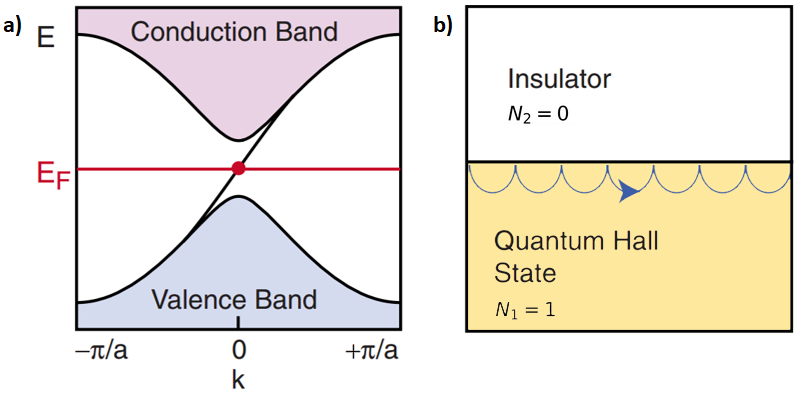}}
\end{minipage} 

   \caption{(a) energy bands of a finite size topological material in contact with an insulator. A branch of edge modes links the two bands. $E_F$ stands for the Fermi level. (b) schema of this system in real space. The blue line shows a propagating chiral edge mode. Figures adapted from \cite{hasan_kane}.
   }
    \label{fig0.1}   
\end{figure} 

Further, according to the fundamental \textit{bulk-edge correspondence theorem}, the number of chiral edge modes at the Fermi energy is equal to $|N_1-N_2|$. This theorem proves the direct topological origin of theses edge modes, which thus inherits from the robustness of the global topological properties of the system, making their propagation particularly robust against spatial disorder and thus to backscattering.

\textbf{Topology and photonics:}

Initially limited to the solid state physics, this topic progressively implemented in other platforms as liquid helium (2006,\cite{volovik}), ultracold atomic gases (2008, \cite{cooper}), and also photonics. Indeed in the latter case, as pointed out by Haldane and Raghu (2008, \cite{haldane}), the physics of light in spatially periodic devices offers large analogies with the physics of electrons in a solid. Since the pioneering observation of chiral edge-states (2009, \cite{soljacic}) in a gyromagnetic material, topological photonics has been a rapidly
growing field. Up to date, different types of platforms have been used to explore topological phases in photonic systems, like coupled resonators (2013, \cite{hafezi}), coupled waveguides (2013, \cite{rechtsman}) or polaritons (2017, \cite{baboux}), and a rich variety of topological phenomena observed. This decade of intense experimental and theoretical efforts has recently led to the release of a Review of Moderns Physics (2019, \cite{ozawa2018topological}) under the impulsion of my internship advisor. 

\textbf{Topological laser:}

\begin{figure}[H]
    \centering
    \includegraphics[scale=0.70]{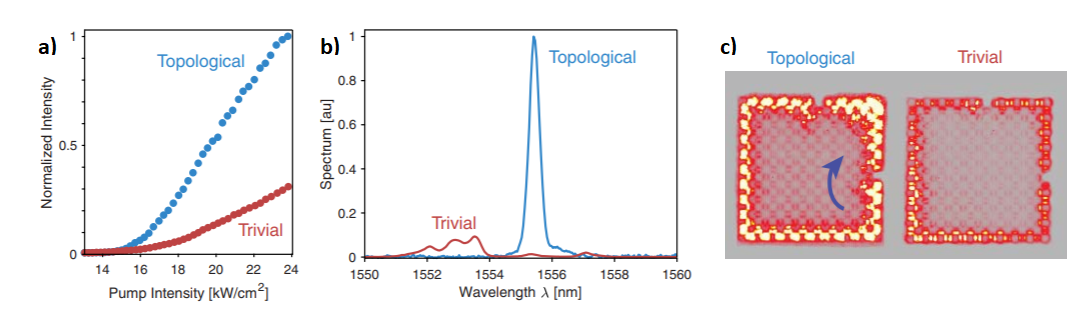}
    \caption{Advantages of a topological laser (blue) versus a trivial (red) one. (a), slope efficiency of the laser emission. (b),  spectrum of the lasing modes. (c), emission of light of the lattice from the top in presence of spatial disorder on the edge. The blue arrow indicates the bypassing of a defect by the chiral edge modes. Figures from \cite{bandres2018topological}.
    }
    \label{fig0.2}
\end{figure}

Recently, new directions of researches focused on the interplay between topological features and intrinsic properties of photonics systems. Among them, a particularly promising device is the so-called \textit{topological laser}. The idea behind it is quite simple: embed a topological lattice in a gain medium tailored so that the lasing mode is an edge mode. After being predicted theoretically (2016, \cite{kibble, conti}), the first experimental demonstration has been made in 2017 \cite{st2017lasing} on a 1D Su-Schrieffer-Heeger lattice of polariton micropillars. It has soon after been demonstrated in a 2D lattice of a time-reversal-breaking photonic crystal embedding magnetic YIG elements (2017, \cite{bahari2017nonreciprocal}), and in a ring resonator array embedding quantum well emitters that provide optical gain (2018, \cite{bandres2018topological}). In this last article, major advantages of 2D topological lasers were pointed out, as we can see in \ref{fig0.2}. 

In all the figures, the comparison of the results is made between the topological lattice and a trivial lattice with nearly the same geometry. (a) shows that the threshold is approximately the same but the slope efficiency is way higher for the topological lattice. (b) shows that the emission of the topological lattice is monomode while the spectrum of the trivial one is broad. Finally, (c) shows that the edge intensity (the optical gain is situated on the last site on the edge on the whole contour) is way more suppressed in the trivial lattice than in the topological one in presence of spatial disorder on the edge.

In spite of these appealing results, little is known about the other properties of 2D topological lasers. In particular, stable emission under continuous pumping of the gain is still not proven, as the experiment in \cite{bahari2017nonreciprocal} is very peculiar and the in the one of \cite{bandres2018topological} the pumping of the gain is operated through short pulses. A recent theoretical work (2018,  \cite{longhi2018presence}) modelling this last experiment claimed that it must not be stable, with an argument of slow career dynamic. Another study (2019, \cite{secli2019theory}) made in our group unveiled also an ultraslow relaxation time of perturbations in such systems. It is eventually in this quasi uncharted land that the topic of my internship takes places, aiming at characterizing in detail the "diagram of phase" of the dynamical stability of 2D topological lasers under the relevant microscopic parameters and studying the elementary excitations of the lasing edge modes to understand the origins of possible instabilities.

To this end, in the first chapter, we will present the driven-dissipative Harper-Hofstadter model, which became the standard model of study (\cite{longhi2018presence, secli2019theory}) since used in the experiment of \cite{bandres2018topological}. In the second chapter, we will present the main results of the two recent articles \cite{longhi2018presence, secli2019theory} dealing with the dynamical stability of 2D topological lasers. In the third chapter, we will present two different methods allowing us to compute the elementary excitations, that we will use in the fourth chapter to shed light on the ultraslow relaxation time of \cite{secli2019theory}, and in the fifth chapter to tackle the problem of non-linearities.

\chapter{General model: topological and driven-dissipative Harper-Hofstadter model}\label{chap1}

In this chapter, we introduce our system of study, and discuss some basic properties of it.

\section{Four bands Harper-Hofstadter lattice}

For our study, we focus on a square Harper-Hofstadter lattice which is known to allow topological lasing \cite{harari2018topological, longhi2018presence, secli2019theory}. The real space Hamiltonian of this model reads in the Landau gauge:

\begin{equation}\label{eq:2.1}
    H =  \sum_{m,n} \left\{  \omega_{ref} \ \hat{a}^{\dagger}_{m,n} \hat{a}_{m,n} - J( \hat{a}^{\dagger}_{m,n} \hat{a}_{m+1,n} + e^{-2\pi i \theta m} \hat{a}^{\dagger}_{m,n} \hat{a}_{m,n+1} + h.c.) \right\}    
\end{equation}

where the sum runs over all the sites of the lattice, $m \in [(0,N_x-1)]$  corresponding to the axis called \textit{x} ($n \in [(0,N_y-1)]$ to the \textit{y} one), $\omega_{ref}$ is the natural frequency of a lattice site, $\hat{a}_{m,n}$ is the photon annihilation operator at the site $(m,n)$ and $J$ a real hopping amplitude. In the following, we will place in the rotating frame at frequency $\omega_{ref}$, so we can drop the first term in (\ref{eq:2.1}). The strength of the synthetic magnetic field is quantified by the flux $\theta$ per plaquette in units of the magnetic flux quantum. For rational $\theta = p/q$, the bulk eigenstates distribute in $q$ bands. For $\theta = 1/4$, there are 4 bands, leading to a Dirac point at $0$ energy and to 2 gaps which exhibit topological properties, associated to opposite Chern numbers $C=\pm 1$ \cite{ozawa2018topological} thus leading to edge modes in each gap for each edge.

\begin{figure}[H]   
\begin{minipage}[t]{0.4\textwidth}
For periodic boundary conditions (PBC) along the \text{y}-axis, the dispersion has the form of Fig.\ref{fig2.1} which shows the chiral edge modes inside the top and bottom gaps. Note that the dispersions of the edge states are shifted into the gap according to the value of $N_x (\ mod \ l_m)$ where $l_m$ is the magnetic length, equals to $1/\theta=4$. This is why we will take the convention $N_x\equiv \ 0 (\ mod \ l_m)$ in the following. 
\end{minipage}
\begin{minipage}[t]{0.6\textwidth}
    \centering
    \raisebox{\dimexpr-\height+\ht\strutbox\relax}{\includegraphics[scale=0.4]{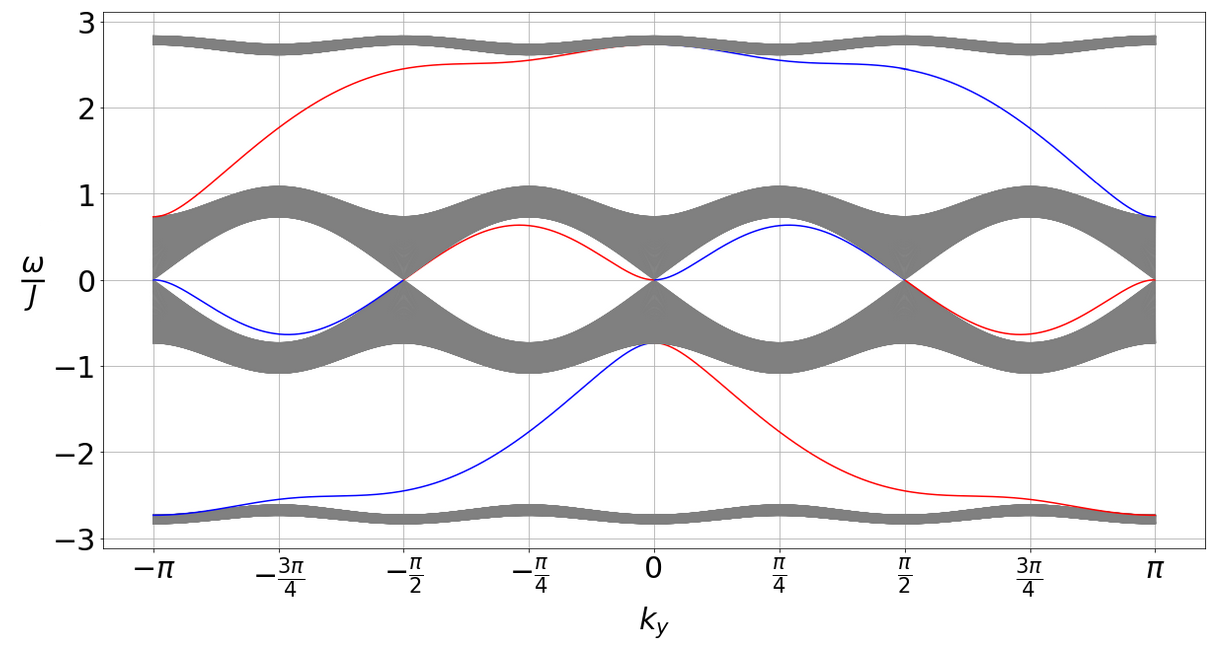}}
 \caption{Energy bands of the Harper-Hofstadter Hamiltonian (\ref{eq:2.1}) with flux $\theta=1/4$ in a lattice of $N_x=399$ sites along $x$ and periodic boundary conditions along $y$. Blue (red) corresponds to the edge state in $x=0$ ($x=398$)}\label{fig2.1}    
\end{minipage}      

\end{figure}

\section{Equations of motion with gain and loss mechanism}\label{chap1sec2}

In the Heisenberg picture, the equation of motion of the photon operator is given by:

\begin{equation}\label{eq:2.2}
    i\frac{\partial \hat{a}_{m,n}}{\partial t} = [\hat{a}_{m,n}, \hat{H}] = -J(\hat{a}_{m+1,n} + \hat{a}_{m-1,n} + e^{-2\pi i \theta m}\hat{a}_{m,n+1} 
      + e^{+2\pi i \theta m}\hat{a}_{m,n-1})
\end{equation}

We now want to add a gain and loss mechanism at the semiclassical level, so we can give up the operator nature of $\hat{a}_{m,n}$ and treat it in a classical way. Then we can introduce a gain and loss mechanism including a reservoir, as well as a non-linear terms \cite{carusotto2013quantum}. Finally the equation describing the time-evolution of the classical field amplitude $\psi_{m,n} $ on each site reads:

\begin{equation}\label{eq:2.3}
     i\frac{\partial \psi_{m,n}(t)}{\partial t} = -J\big[\psi_{m+1,n} + \psi_{m-1,n} + e^{-2\pi i \theta m}\psi_{m,n+1} 
      + e^{+2\pi i \theta m}\psi_{m,n-1}\big] + \big[ g|\psi_{m,n}|^2  + g_R N_{m,n} + i(R N_{m,n} - \gamma) \big] \psi_{m,n}    
\end{equation}

where the reservoir density $N_{m,n}$ is determined by the rate equation

\begin{equation}\label{eq:2.4}
      \frac{\partial N_{m,n}}{\partial t} = P_{m,n} - (\gamma_R + R|\psi_{m,n}|^2)N_{m,n} 
\end{equation}

where $g$ is the strength of the non-linear interaction between 2 particles in the lasing mode, $g_R$ is the strength of the non-linear interaction between a particle in the lasing mode and one in the reservoir, $R$ is the amplification term, $\gamma$ is the on-site loss rate, $\gamma_R$ is the reservoir loss rate and $P_{m,n}$ is the pump intensity at the site $(m,n)$.

Nb: note that the addition of terms in this way changes the spectrum (energy bands) of the system compared to the case of the bare HH model. This change can be neglected for low $\gamma$ and $P$ values, but can also have major consequences as discussed in Chapter \ref{chap5}.

Finally, we can focus on two particular cases that will be of particular interest in the following. The first consists in a system with periodic boundary conditions along the y-axis, and a homogeneous pumping along this direction on one edge ($P_{m,n} = P_m \delta_{m,0}$). In this case, the system is translationnary invariant along the $y$-axis, so we can apply the Bloch theorem. The result is  that a steady-state of the system (when existing) can be written on the form $\psi^{SS}_{m,n}(t) = \psi^0_m e^{-i(\mu t -k_{y,0} n)} $ and $N^{SS}_{m,n}= N^0_m$ where $\mu$ is the energy and $k_{y,0}$ the wavevector of the steady state. This simplifies a lot the dependence of the steady-state in the two spatial variables.

The second consists in doing the so-called adiabatic approximation for the reservoir \cite{wouters2007excitations}. It becomes possible when the dynamic of the reservoir (ruled by the time scale $1/\gamma_R$) is much faster than the dynamic of the lattice (time scale $1/\gamma$), i.e. $\gamma_R / \gamma \gg 1$. In this case, the reservoir can be considered as being in its stationary regime at any time ($\frac{\partial N_{m,n}}{\partial t} = 0 \ \forall t$), so injecting (\ref{eq:2.4}) into (\ref{eq:2.3}), we get:

\begin{equation}\label{eq:2.5}
     i\frac{\partial \psi_{m,n}(t)}{\partial t} = -J\Big[\psi_{m+1,n} + \psi_{m-1,n} + e^{-2\pi i \theta m}\psi_{m,n+1} 
      + e^{+2\pi i \theta m}\psi_{m,n-1}\Big] + \Big[ g|\psi_{m,n}|^2 + i(\frac{\chi P^{ad}_{m,n}}{1+\beta |\psi_{m,n}|^2} - \gamma) \Big] \psi_{m,n}    
\end{equation}

where $\beta= \frac{R}{\gamma_R}$, $P^{ad}_{m,n} = \beta P_{m,n}$ and $\chi = (1-i\frac{g_R}{R})$. 

This simplifies a lot the problem suppressing $N_x \times N_y$ equations.

\section{Lasing threshold}\label{chap1sec3}

We want to investigate the lasing properties of the HH lattice that we defined previously. So one first information that we need is the lasing threshold whatever is the geometry of the system.

The lasing threshold is defined as the pumping power above which the trivial ($\psi_{m,n} = 0 \  \forall (m,n)$) solution becomes unstable. We then perform a linearization of (\ref{eq:2.4}) around this trivial solution, taking the ansatz $\psi_{m,n}(t) = \delta \psi_{m,n} e^{-i\omega t}$ with $|\delta \psi_{m,n}| << 1$. We obtain that $N_{m,n} = \frac{P_{m,n}}{\gamma_R} + o(|\delta \psi_{m,n}|^2)$. Thus, injecting this result in (\ref{eq:2.3}), we get the equation:

\begin{equation}\label{eq:2.5}
     -J\big[\delta \psi_{m+1,n} + \delta \psi_{m-1,n} + e^{-2\pi i \theta m} 
     \delta \psi_{m,n+1} 
      + e^{+2\pi i \theta m}\delta \psi_{m,n-1}\big] + i\big[\chi P^{ad}_{m,n} - \gamma\big] \delta \psi_{m,n} = \omega \delta \psi_{m,n}    
\end{equation}
with $P^{ad}_{m,n} = \frac{R P_{m,n}}{\gamma_R}$ and $\chi = (1-i\frac{g_R}{R})$.

Putting this equation in a matricial form, this system can be solved easily with a computer, with all type of boundary conditions.

The real part of the eigenvalues $\omega$ of the matrix gives us access to the value of the frequency of the modes that can perturb the trivial solution. When the imaginary part of a given $\omega$ becomes $\ge 0$, then the perturbation grows when propagating in the lattice, which makes the system unstable. Thus we can find the lasing threshold by tuning the pump intensity from low values until one eigenvalue reach the $Im(\omega)=0$ threshold.  
  
\begin{figure}[H]   
\begin{minipage}[t]{0.4\textwidth}
 A typical example is shown in Fig. \ref{fig2.2}, for a lattice with periodic boundary condition along the $y$-axis and pumping along the $x=0$ edge (i.e. $P^{ad}_{m,n} = P^{ad} \delta_{m,0}$ where $\delta_{a,b}$ is the Dirac function). The essential idea to keep in mind is that the system will tend to lase in the middle of the gap, where the gain is maximal. Also, we see that the gain distribution is the same in the two gaps, which means that lasing can start randomly with same probability in the two gaps. Finally, note that the allowed frequency for lasing are quantized due to the periodic boundary conditions. The spacing between the allowed modes is thus bigger as $N_y$ gets smaller.
\end{minipage}
\begin{minipage}[t]{0.6\textwidth}
    \centering
    \raisebox{\dimexpr-\height+\ht\strutbox\relax}{\includegraphics[scale=0.6]{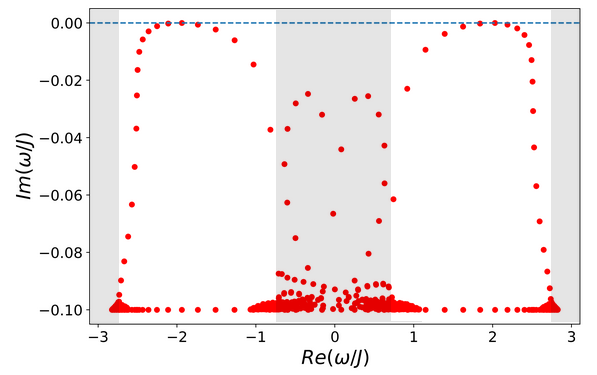}}
    \caption{Imaginary part of the elementary excitation spectrum on top of the trivial solution as a function of the frequency. The white (grey) background indicates the gaps (bands). Parameters: $\gamma /J = 0.1$, $P^{ad}_{th}= 1.142\gamma$, $N_x=31$, $N_y=41$, periodic boundary conditions along $y$}
    \label{fig2.2}    
\end{minipage}      
\end{figure}

\section{Time evolution computation}

Now we know the threshold of emission of the system, we want to go further in simulating the time evolution of the system, according to (\ref{eq:2.3}),(\ref{eq:2.4}). 

The algorithm to compute the time evolution is the one developed by Matteo Secli, a former student of the group, and corresponds to the algorithm used in \cite{secli2019theory}. It consists in a fourth-order Runge-Kutta algorithm (global truncation error $O(h^4)$ where $h$ is the integration step), which prevents deviations from the physical solution due to discretization. The algorithm can be started in two ways: one being a small complex random noise on every lattice site; the other being a specific spatial pattern to favour lasing in a given mode (for example the function $\psi^{seed}_{m,n}= e^{ik_{y,0}n}$ for a system with periodic boundary conditions along the $y$-axis and pumped along one edge will trigger lasing in the edge mode with wavevector $k_y,0$). 

The results from the simulation can be twofold: either the system converges to a steady-state (SS), namely reaches a regime with no time evolution of the intensity in every site, so we will say the system is dynamically stable; either even after a long time the intensity on some sites is still modulated in time in a non-converging way and we will say that the system is dynamically unstable. When the system is stable, we say that the system reached the SS when the fluctuations of the intensity in time are of the order of $10^{-13}$, which corresponds to the limit of the numerical precision. 

When a SS is reached, we let it evolve and then compute the Fourier transform of the field amplitude on all the pumped sites. Eventually we do the mean between all the sites to get a precise estimate of the lasing frequency $\mu$. When the system is invariant along one direction (PBC and homogeneous pumping along this direction), we can extract the wavevector by computing the spatial Fourier transform along the periodic direction for a strip (usually the strip which is pumped). This process can be repeated on several times to make an average.

Finally, using a computed has a major advantage compared to a real system to get knowledge about unstable phases. Indeed, when letting evolve the system in a stable phase to reach a SS, and then changing the parameters suddenly to go to an unstable phase, the system in general has a quasi-stable evolution in the unstable phase, the time for very small numeric perturbations to grow and make the state explode. During this quasi-stable evolution, it allows us to use the state of the system to compute further properties of the system, and in particular the elementary excitations spectrums, as we will see in the next chapters.

\chapter{Dynamical stability of the driven-dissipative Harper-Hofstadter model: state of the art}\label{chap2}

In this chapter, we present the two first articles which dealt with dynamical stability of 2D topological lasers, and which are both relying on the driven-dissipative HH model. 

\section{\textit{Theory of chiral edge state lasing in a two-dimensional topological system }(2019) \cite{secli2019theory}}\label{chap2sec1}

This article has been written recently by the former student of the group Matteo Secli, and is the starting point of the investigation we lead during this internship. We will summarize here the main results which count for our work.

\textbf{System:}

The system of study in this article is a square Harper-Hofstadter lattice with flux $\theta=1/4$. The pump is taken in the adiabatic approximation, with no non-linear interactions. Thus the time-evolution equations of the classical field amplitude read:

 \begin{equation}
     i\frac{\partial \psi_{m,n}(t)}{\partial t} = -J\Bigg[\psi_{m+1,n} + \psi_{m-1,n} + e^{-2\pi i \theta m}\psi_{m,n+1} 
      + e^{+2\pi i \theta m}\psi_{m,n-1}\Bigg] + i\Bigg[\frac{P^{eff}}{1+\beta |\psi_{m,n}|^2} - \gamma  \Bigg] \psi_{m,n}
\end{equation}

where the parameters are the same as defined in Chapter \ref{chap1} Section \ref{chap1sec2}.

\textbf{Good pumping scheme:}

\begin{figure}[H]   
\begin{minipage}[t]{0.5\textwidth}
The first experimental demonstration of a 2D topological laser \cite{bahari2017nonreciprocal} has been made through whole system pumping (WSG), i.e. $P_{m,n} = P \ \forall (m,n)$ (thought on a different type of lattice). Thus it is natural to wonder if it is possible to trigger topological lasing in this model through whole system pumping too. In the article, simulations were done to investigate it, and the result is shown in Fig. \ref{fig22.1} (a). We see that the laser emission is spread throughout the whole system, and time resolved simulations showed that such a spatial modulation persists indefinitely in a chaotic manner. This means that no monochromatically oscillating steady state is ever reached, a fortiori the lasing is not topological.

\end{minipage}
\begin{minipage}[t]{0.5\textwidth}
    \centering
    \raisebox{\dimexpr-\height+\ht\strutbox\relax}{\includegraphics[scale=1.]{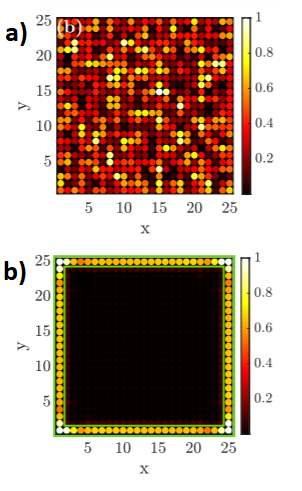}}
  
\end{minipage} 
    \caption{ Topological lasing in a $25 \times 25$ HH lattice with $\theta=1/4$ for $\beta=1$ and $\gamma = 0.1J$. (a) Snapshot of the (normalized) intensity distribution at $t=1000/\gamma$ for WSG. (d) Same for a one-site-thick WEG. The green rectangle indicates the amplified sites. Figures from \cite{secli2019theory}.}
    \label{fig22.1}  
\end{figure}

The satisfying pumping scheme to favour lasing in an edge state was rather found to be pumping only the edge, when the overlap between the spatial position of the edge modes and the pump profile is maximal. As we can see in Fig. \ref{fig22.1} (b), when pumping the whole edge (Whole Edge Gain), a monochromatical steady-state is reached with intensity localized on the edge of the lattice and frequency inside a gap (not shown) evidencing for its topological nature. Also, as the gain is broadband, the probability to start lasing in one gap or in the other is the same (as in Fig. \ref{fig2.2} for a system with PBC). 

\textbf{Ultraslow relaxation time}

A notable feature was found in the favorable WEG configuration: when the system is let free to evolve to a steady-state after an initial random noise, a lasing steady-state in an edge mode is quickly obtained, but on top of it a small perturbation is still present and is damped on a time scale longer than any time scale in the system, namely $\gamma$ or $\frac{L}{v_g}$(which corresponds to the approximate round-trip time around the cavity, where $L$ is the perimeter of the cavity and $v_g$ the group velocity of the edge mode).

\begin{figure}[H]   
\begin{minipage}[t]{0.4\textwidth}
This phenomena is illustrated in Fig. \ref{fig22.2}, where we can see the evolution of the intensity profile on one side of the square for increasing time (normalized to a round trip, which is longer than a time step). This clearly shows that the relaxation of the perturbation is very slow, since 5 round trips only divide the initial amplitude by 3.
\end{minipage}
\begin{minipage}[t]{0.6\textwidth}
    \centering
    \raisebox{\dimexpr-\height+\ht\strutbox\relax}{\includegraphics[scale=1.2]{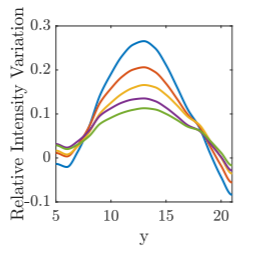}}
 
\end{minipage}
   \caption{Cuts of the intensity distribution along the bottom side of the square at times (from top to bottom) $\gamma t = 43.85,\ 51.65,\ 59.45,\ 67.20,\ 75.00 $. Figure from \cite{secli2019theory}.
    }
   \label{fig22.2}    
\end{figure}

It is this first intriguing result that called for building a theory of small excitations around a lasing topological edge mode, as it will be developed in the next chapter.

\section{\textit{Presence of temporal dynamical instabilities in topological insulator lasers} (2018) \cite{longhi2018presence}}\label{chap2sec2}

This article is the first one which focused on the dynamical stability of 2D topological lasers. More precisely, its goal was to show that under continuous wave pumping, a system such as the one in \cite{bandres2018topological} should not be stable. We will present this key point of the article in the following of this section.

\textbf{System:}

The system of study in the article is a square Harper-Hofstadter lattice with flux $\theta = 1/4$, and with pumping on the whole contour of the lattice through a reservoir. The main difference with (\ref{eq:2.3}) and (\ref{eq:2.4}) is that the loss rate is not the same in the bulk as on the edge, and that there is an output coupler in one corner of the lattice. The time evolution of the classical field amplitude $\psi_{m,n}$ in this system is:

\begin{equation}\label{eq:3.8}
\begin{split}
        i\frac{\partial \psi_{m,n}(t)}{\partial t} = -J[\psi_{m+1,n} &+     \psi_{m-1,n} + e^{-2\pi i \theta m}\psi_{m,n+1} + e^{+2\pi i     \theta m}\psi_{m,n-1}] + g_R N_{m,n} \psi_{m,n} \mathbb{1}_{(m,n) \in  \mathcal{P}}   \\
     &+  i (N_{m,n}-\gamma_p ) \psi_{m,n}\mathbb{1}_{(m,n) \in  \mathcal{P}}   
     - \gamma \psi_{m,n} \mathbb{1}_{(m,n) \notin  \mathcal{P}}
     - \gamma_{out} \psi_{0,0} \mathbb{1}_{(m,n)= (0,0)}
\end{split}
\end{equation}

where the reservoir density $N_{m,n}$ is defined only on the contour ($(m,n) \in \mathcal{P}$) and is determined by the rate equation:

\begin{equation}
      \frac{\partial N_{m,n}}{\partial t} = P - \gamma_R (1 + |\psi_{m,n}|^2)N_{m,n} 
\end{equation}

where $J$ is the coupling constant between the sites,  $\gamma_p$ is the loss rate on the edge, $g_R$ is the linewidth enhancement factor (namely an interaction between the particles from the lasing mode and from the reservoir), $\gamma$ is the loss rate in the bulk, $\gamma_{out}$ is the loss rate due to the output coupler, $P$ is the pump intensity, $\gamma_R$ is the loss rate in the reservoir, $\mathbb{1}$ the indicator function and $\mathcal{P}$ indicates the perimeter of the lattice.

\textbf{Influence of the reservoir speed:}

\begin{figure}[H]   
    \centering
\includegraphics[scale=0.85]{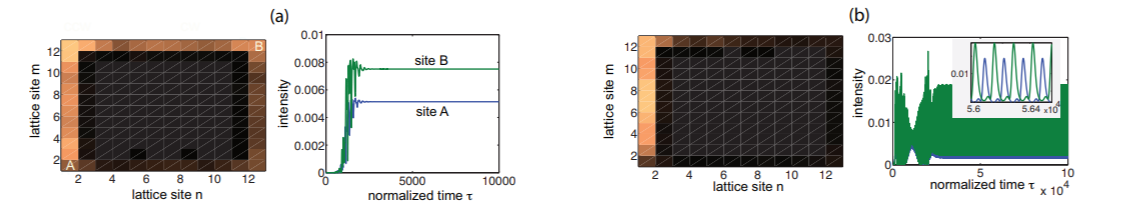}

   \caption{ Transient switch-on dynamics of the topological insulator laser made of $12 \times 12$ sites. Time is normalized to the cavity photon lifetime $\frac{1}{\gamma_p}$ (i.e. $\tau = t\gamma_p$). Parameter values are: $\theta= 1/4$, $J= 5 ns^{-1}$,  $g_R=-3$, $\gamma_p= 2.5J$, $\gamma= 0.5J$, $\gamma_{out}= 1J$, $P/\gamma_R= 1.0392 \gamma_p$ and  $\gamma_R= 10\gamma_p$ in (a) (virtual limit of class-A
laser), $\gamma_R= 0.02 \gamma_p$ in (b) (class-B laser).
The pump rate $P/\gamma_R= 1.0392 \gamma_p$ is chosen very close to its threshold value. Left panels show the normalized intensity distribution $|\psi_{m,n}|^2$ in the lattice at final time [$t= 0.5\times 10^4/\gamma_p = 0.4 \mu s$ in (a) and $t= 0.5\times 10^5/\gamma_p = 4 \mu s$ in (b)]. Right panels show the temporal evolution of the field intensity at the vertex sites A (output coupler) and B. Adapted from \cite{longhi2018presence}}
    \label{fig0.3}
\end{figure}

The main result of this article is that for relevant experimental parameters (for \cite{bandres2018topological}), starting from a small random noise, the lasing on the edge is stable when the ratio $\frac{\gamma_R}{\gamma_p} \gg 1$, and becomes unstable when  $\frac{\gamma_R}{\gamma_p} \ll 1$, as can be seen in Fig. \ref{fig0.3}.

This kind of transition from stable to unstable when lowering the speed of the reservoir is a well-known fact in laser \cite{baili2009direct} and polariton condensates \cite{wouters2007excitations, baboux2018unstable} physics. This way, it has to be understood more deeply if topology can help to protect this new kind of lasers from dynamical instabilities or not. This is what we will try to answer in the following work.

\chapter{Elementary excitations of a lasing chiral edge mode on a cylinder}\label{chap3}

The purpose of our work is to shed light on the dynamical stability of topological lasers through their elementary excitations. To this end, we introduce in this chapter the several models that we will use to compute them, starting from a precise 2D model, then deriving an effective 1D model simpler to manipulate. 

\section{2D cylindrical model}

The first articles presented in the last chapter use a square Harper-Hofstadter lattice pumped on the whole edge. However this geometry is not the most practical to handle. Indeed, we want to compute the elementary excitations spectrum of the system, which implies linearizing the set of equations (\ref{eq:2.3}),(\ref{eq:2.4}) and then diagonalizing the corresponding Bogoliubov matrix of size $3\times N_x \times N_y$, which is computationally demanding when the lattice is big. This is why a cylindrical geometry (namely Periodic Boundary Condition along one axis) is much more practical, since under homogeneous pumping the translational invariance allows to pass into Fourier space, thus making the problem of size $3\times N_x$.
Furthermore, we claim that for a big lattice, the square and the cylindrical geometry are equivalent. Indeed, a chiral edge mode propagates undirectionnaly on the edge and is not affected by little geometrical perturbations thanks to topology, so a square can be mapped to a circle safely. This way, all the results we derive in this chapter and the next ones directly apply to these articles.

Finally, we can recall the equations of motion of the system in this geometry, choosing the $y$-axis to be periodic and pumping homogeneously along the $x=0$ edge:

\begin{equation}\label{eq:4.1}
    \left \{
\begin{array}{l}
\begin{split}
i\frac{\partial \psi_{m,n}}{\partial t} = -J[\psi_{m+1,n} + \psi_{m-1,n} &+ e^{-2\pi i \theta m}\psi_{m,n+1} 
      + e^{+2\pi i \theta m}\psi_{m,n-1}] \\
      & + [ g|\psi_{m,n}|^2 + g_R N_{m,n} + i(R N_{m,n} - \gamma) ] \psi_{m,n}   
\end{split}\\ 
\frac{\partial N_{m,n}}{\partial t}(t) = P \delta_{m,0} - (\gamma_R + R|\psi_{m,n}|^2) N_{m,n}
\end{array}
\right.
\end{equation}

where $\psi_{m,N_y} = \psi_{m,0} \ \forall m$ and $N_{m,N_y} = N_{m,0} \ \forall m$.

\section{Linearization of the driven-dissipative Harper-Hofstadter model}

We now want to linearize our 2D model HH in order to compute the elementary excitation spectrum. 

We know that the steady-state of these equations can be written as $\psi^{SS}_{m,n}(t) = \psi^0_m e^{-i(\mu t -k_{y,0} n)} $ and $N^{SS}_{m,n}(t)= N^0_m$ so we linearize taking the ansatz: $\psi_{m,n}(t) = (\psi^0_m + \delta \psi_{m,n}(t))e^{-i(\mu t -k_{y,0} n)} $ and $N_{m,n}(t)= (N^0_m + \delta N_{m,n}(t))$, with $|\delta \psi_{m,n}| \ll |\psi^0_{m}|$ and $\delta N_{m,n} \ll N^0_{m}$. Note that $\delta \psi_{m,n}$ is complex, so is independent from $\delta \psi_{m,n} ^*$, which will follow an independent equation of evolution.

Doing straightforward calculations and passing to Fourier space thanks  to translational invariance along the $y$-axis, we get the system:

\begin{equation}\label{eq:3.333}
      \left \{
\begin{array}{l}
\begin{split}
    i\frac{\partial \delta \psi_{m,k}}{\partial t} = -J[\delta \psi_{m+1,k} +  \delta \psi_{m-1,k} + 2cos(2\pi \theta m +  k_{y,0} + k) \delta \psi_{m,k}] + g\psi^0_m \ ^2 \delta \psi^*_{m,-k} \\
    +[2g|\psi^0_m|^2 + g_R N_m^0 - \mu + i(R N^0_m - \gamma)  ]\delta \psi_{m,k} + [g_R \psi_m^0 + iR \psi^0_m ]\delta N_{m,k}
\end{split}\\ 
      \frac{\partial \delta N_{m,k}}{\partial t} = - (\gamma_R + R|\psi^{0}_{m}|^2)\delta N_{m,k} + iR N^{0}_{m} ( \psi^{0}_{m} \delta \psi^{*}_{m,-k} + \psi^{0*}_m \delta \psi_{m,k}) 
\end{array}
\right.
\end{equation}

This problem can be put into a matricial form and solved numerically for each $k$ in an easy way, giving the elementary excitations spectrum of the system.

Nb: the calculation to compute the elementary excitations in the adiabatic approximation is similar to this one and is not showed here.

\section{Parameters and spectrum of reference}\label{secref}

\textbf{Parameters of reference:}

Before investigating the elementary excitations spectrum with the previous method, we choose a set of typical parameters that will serve in all the following:

\begin{itemize}
    \item We will express all our results in units of the hoping amplitude $J$.
    
    \item In order to make the situation of reference the most simple possible, we place in the adiabatic approximation, i.e. $\gamma_R/\gamma = + \infty$ and with no non-linear term, i.e. $g,\ g_R = 0$ ($\implies \chi=1$).
    
    \item We take $\gamma = 0.1 J$, which means that the losses are relatively small compared to the energy of the lattice, so corresponds to negligible modifications of the Harper-Hofstadter energy spectrum due to the non-hermitian terms.
    
    \item We express the pump in unit of $\gamma$, as it corresponds to the lasing threshold of an isolated site. For a lattice with $N_y= 41$ sites along the $y$-axis, the laser threshold is situated at $P^{ad}_{th}= 1.142\gamma$, and corresponds to the mode \{ $k_{y,0}= -0.9195$, $\mu/J = -1.9374$ \} in the lower gap and \{ $k_{y,0}= 2.1455$, $\mu/J = 2.0285$ \} in the upper gap. (Note that the energies in the two gaps are not exactly the same due to the discretization of the modes, see Fig. \ref{fig2.2}).  
    
    \item In practice, to make the comparison easier between different results, we will put a seed in the initial condition to trigger lasing in the \{ $k_{y,0}= -0.9195$, $\mu/J = -1.9374$ \} mode.
    
    \item The value $P^{ad}=1.25\gamma\ = 1.095 P^{ad}_{th}$ is a good compromise between a high enough gain to start lasing quickly and low enough gain to keep the non-hermitian contribution negligible. 
    
    \item We take $\beta = 0.01$, even if in practice the results are independent of this parameter (see section \ref{chap3sec7}.
    
    \item We take the size $N_x= 31$, much bigger than the penetration length of the edge modes in the bulk to avoid finite size problems.

\end{itemize}

These parameters correspond to the parameters of reference. They are the one that will be used in the following. Any modification of one of these parameters will be explicitly written.

\newpage
\textbf{Spectrum of reference:}

Taking the previous parameters, we can make a first simulation of the system. This set of parameters leads to a steady-state, so is in the stable phase. We extract from the steady state a given layer along the $x$-axis (the SS is invariant along the $y$-axis) that we inject in the Bogoliubov matrix from the (\ref{eq:3.333}) system.

\begin{figure}[H]
    \centering
    \includegraphics[scale=0.85]{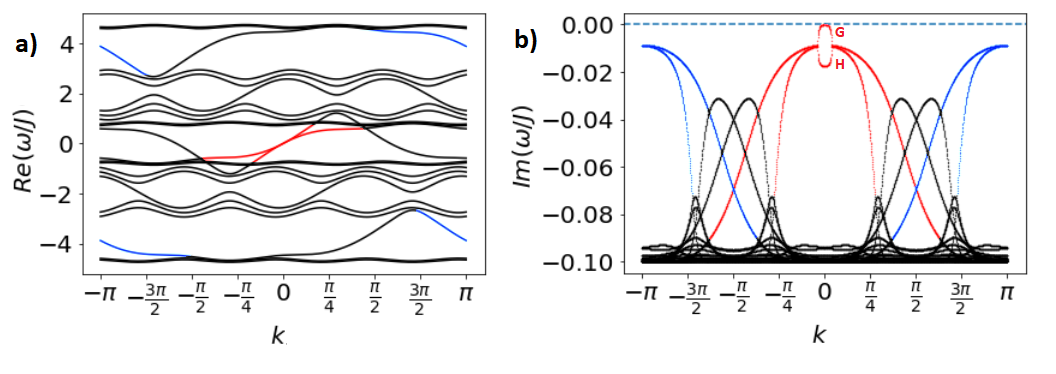}
    \caption{Real (a) and imaginary (b) part of the elementary excitations spectrum as a function of the wavevector $k= k_y- k_{y,0}$. The red (blue) curves in (a),(b) correspond to the elementary excitations around the lasing chiral edge mode (the edge mode on the same edge with reverse chirality). The black curves correspond to trivial modes.
    }
    \label{fig3.1}
\end{figure}

The result is given in Fig. \ref{fig3.1}. In this figure, for the real and imaginary part, the red part corresponds to the elementary excitations around the lasing edge mode. The blue part corresponds to the edge modes on the same edge but with reverse chirality. The remaining black lines corresponds to trivial modes. 

The first remarkable features are the following. In the real part of the spectrum, the excitations around the lasing mode exhibit a zone of adhesion, as it is typical in non-equilibrium condensates \cite{wouters2007excitations,baboux2018unstable, chiocchetta2013non}. The main difference with these later is the slope of the branch, due to the non-zero group velocity of an edge mode (and further odd-derivative terms, see section \ref{chap3sec7}). 

In the imaginary part of the spectrum, we recognize again the typical splitting between Goldstone (G) and Higgs (H) modes (in red at the center of the figure)  around the lasing mode. We naturally find the $Im(\omega/J)=0$ value of the Goldstone mode at $k=0$ consequence of the Goldstone theorem. Also, we see that all the imaginary part of the spectrum is negative, confirming that the system is stable.

Finally, we note that in this case all the information about the dynamic of the system is concentrated around the lasing mode (red part), through the shape of the Goldstone mode. This statement is of major importance, since it motivated us in developing an effective model able to capture this relevant information in a simple way, as we will see in the next section.

\section{1D effective model}\label{chap3sec7}

In this section, we present a very simple model that is able to reproduce the behaviour at small $|k|$ of the central part of Fig. \ref{fig3.1}. This model is based on a the assumption that as the field is concentrated on the gainy boundary of the lattice, it is enough to focus on the evolution of the $\psi_{0,n}(t)$ function to retrieve the dynamics of the lasing mode. The derivation of this equation can be found in Appendix \ref{appendixA}.

In these approximations, we come to a generalized Gross-Pitaevskii Equation (gGPE) \cite{carusotto2013quantum} for the evolution of the field amplitude:

\begin{equation}\label{eq:3.4}
    i\frac{\partial \psi_n}{\partial t} = \Bigg[ TF^{-1}[\epsilon(k_y)](n)\circledast + g G_0 |\psi_n|^2 + g_R G_0 N_n + i\Big(R G_0 N_n - \gamma\Big)\Bigg]\psi_n
\end{equation}

where the reservoir follows the equation,

\begin{equation}
    \frac{\partial N_n}{\partial t} =
    P - \gamma_R (1+R |\psi_n|^2) N_n
\end{equation}

where $\psi_{m,N_y} = \psi_{m,0} \ \forall m$, $N_{m,N_y} = N_{m,0} \ \forall m$, $\epsilon(k_y)$ is the dispersion of the edge modes of a given gap and $G_0= \frac{\gamma}{P^{ad}_{th}}$ where $P^{ad}_{th}$ has to be calculated as in Chapter \ref{chap1} section \ref{chap1sec3}.

These equations are very close from the ones describing non-equilibrium condensates in a planar cavity \cite{wouters2007excitations, carusotto2013quantum}. It allows to compute numerically the elementary excitation spectrums in an easy way, without requiring to compute a SS.

Further, when the adiabatic approximation is feasible, we can even get analytical results. Indeed, the equation of motion in the adiabatic approximation reads:

\begin{equation}\label{eq:3.5}
    i\frac{\partial \psi_n}{\partial t} = \Bigg[ TF^{-1}[\epsilon(k_y)](n)\circledast + g G_0 |\psi_n|^2 + i\Big(\frac{\chi G_0 P^{ad} }{1+\beta |\psi_n|^2} - \gamma\Big)\Bigg]\psi_n  
\end{equation}

where $\beta = \frac{R}{\gamma_R}$,  $P^{ad} = \beta P$  and $\chi = (1-i\frac{g_R}{R})$.

This time, the density uniform steady-state solution reads $\psi_n(t) = \psi^0 e^{-i\mu t}e^{i k_{y,0}n}$ where $\psi^0 = \sqrt{\frac{1}{\beta}(\frac{P^{ad}}{P^{ad}_{th}(k_{y,0})}-1)}$ and $\mu = \epsilon(k_{y,0}) + g G_0|\psi^0|^2 + \frac{g_R}{R}\frac{G_0 P^{ad}}{1+\beta |\psi_0|^2}$. Using the ansatz $\psi_n(t) = \psi^0(1 + \delta \psi_n(t)) e^{-i\mu t}e^{i k_{y,0}n}$ with $|\delta \psi_n| \ll 1$ and passing to Fourier space, we get the matricial equation:

\begin{equation}\label{eq:3.5}
i\frac{\partial }{\partial t}       
\begin{pmatrix} 
        \delta\psi_k \\ 
        \delta\psi_{-k}^{\ast} \\
      \end{pmatrix} 
      = \\
      \Bigg[
    \epsilon^O(k)\mathbb{1}
     +
         \begin{pmatrix}
    \epsilon^E(k) - i\Gamma + \mu_{tot} & \mu_{tot} - i\Gamma \\
    -\mu_{tot} - i\Gamma & -\epsilon^E(k) -i\Gamma - \mu_{tot}
    \end{pmatrix}\Bigg]
    \begin{pmatrix} 
        \delta\psi_k \\ 
        \delta\psi_{-k}^{\ast} \\
      \end{pmatrix} 
\end{equation}

where $\Gamma = \gamma(1-\frac{P_{th}}{P})$, $\mu_{tot} = g G_0|\psi^0|^2 - \frac{g_R G_0 \Gamma}{R}$ and $\epsilon^E$ ($\epsilon^O$) the even (odd) part of $\epsilon$ (without the order zero $\epsilon(k_{y,0})$).

In this form, we recognize the second term of the matrix to be the same as in \cite{chiocchetta2013non}, thus the double branched excitation spectrum reads: 

\begin{equation}\label{eq:3.7}
  \omega_{\pm} = \epsilon^O(k) - i\Gamma \pm \sqrt{\epsilon^E(k)[\epsilon^E(k) + 2\mu_{tot}] - \Gamma^2} 
\end{equation}

In practice, we can do a last simplification of the problem, only keeping the third first orders of $\epsilon(k) \approx a_1 k + a_2 k^2 + a_3 k^3$.

\begin{figure}[H]
    \centering
    \includegraphics[scale=0.7]{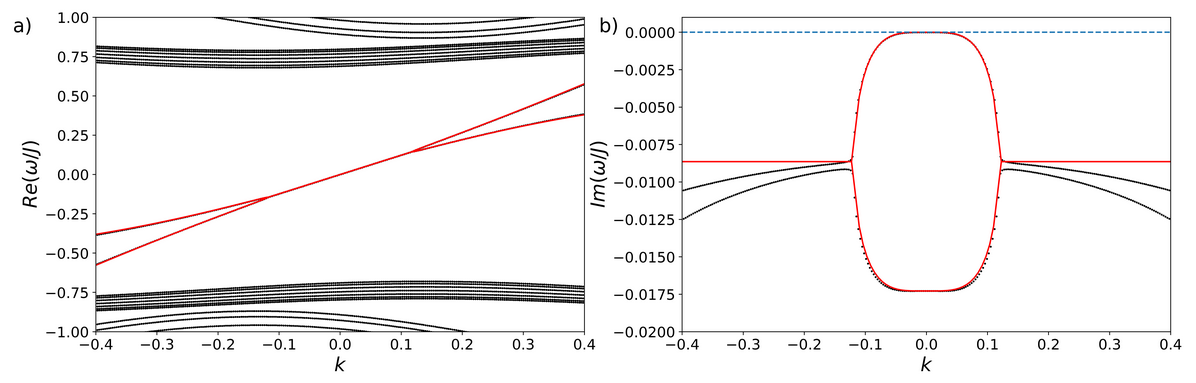}
    \caption{Real (a) and imaginary (b) part of the elementary excitations spectrum for the 2D model (black dots) and the gGPE model (red lines). Parameters of the gGPE model: $G_0 = 0.8754$, $a_1= 1.2310 J$, $a_2= 0.6179 J$, $a_3= -0.2138 J$
    }
    \label{fig3.5}
\end{figure}

 A comparison of this very simple model and the 2D model is showed in Fig. \ref{fig3.5}. We see that the shape of the Goldstone mode is very well reproduced by this model, which is the essential feature that we wanted. The side branches are artefacts due to the brutal approximations made (see Appendix \ref{appendixA}), but are not relevant as long as the lasing point is in one of the maximal regions of Fig. \ref{fig2.2} (in real plots they are supposed to decay to low imaginary values, which is not interesting for the stability analysis). Further comparisons between this model and the 2D one are given in Appendix \ref{appendixB}, and show that this model remains relevant in all the regimes, provided lasing at the maximum of the gain. %The main fact is that this model remains valid as long as the size of the sticking zone (in the real part of the spectrum) between the particle and antiparticle branches remains smaller than the size of the gap, namely roughly $\Gamma < \frac{a_2 a_1^2 \Delta \omega_{gap}}{4}$ where $\Delta \omega_{gap}$ is the energy width of the gap.

As it is analytical, this model thus represents a good model to understand the reasons behind some behaviour of the system. A key remark has also to be done: at the third order in $\epsilon$, the imaginary part of (\ref{eq:3.7}) is the same as in the case of non-equilibrium condensates in a planar cavity \cite{wouters2007excitations, carusotto2013quantum}. This means that all the results that have been established in this kind of systems with this model can be revisited in the case of topological lasers. Some insights will be given in the final conclusion, however these studies are out of the scope of this report.

\section{Conclusion:}

In this chapter, we showed how to calculate the elementary excitations of a lasing chiral edge mode on a cylinder. We started with the direct linearization of the driven-dissipative Harper-Hofstadter model introduced in Chapter \ref{chap1}. This allows a detailed computation of the elementary excitations, and exhibited in a first reference example that the dynamic of the system is ruled by the classic features of non-equilibrium condensates. This statement motivated us to introduce a simple 1D effective model based on a generalized Gross-Pitaevskii Equation for a topological edge mode, and we showed that the elementary excitations spectrum from this model reproduces correctly the shape of the Goldstone mode. The advantages of this model are fourfold: first it allows for calculating the elementary excitation spectrums without computing the SS of the system; second, it allows for analytical calculations in the adiabatic approximation; finally, it is nearly model independent since it only depends on the dispersion of the edge modes.

\chapter{Investigation of the ultraslow relaxation time of a perturbation}\label{chap4}

In the two following chapters, we want to use the models introduced in the last chapter to make a detailed description and interpretation of the dynamical stability of lasing topological edge modes. To this end, we will start in this chapter by explaining the ultraslow relaxation time in \cite{secli2019theory} and then exploring the sets of parameters in which this phenomenon is kept.

\section{Ultraslow relaxation time in Secli and al.}

To start, we remark that the parameters that we took as point of reference are very close to the ones used in \cite{secli2019theory}. As we saw in Chapter \ref{chap2} Section \ref{chap2sec1}, the convergence to the SS of the system is made very slow by the presence of perturbations that keeps on oscillating on a very long time compared to the round-trip time of the lasing edge mode and a fortiori to $\frac{1}{\gamma}$. This can be explained using the elementary excitation spectrum of Fig. \ref{fig3.1}. Indeed, looking at the behaviour of the Goldstone mode around $k=0$, we see that it is very flat in this region. Further, a limited development of (\ref{eq:3.7}) shows that the small $k$ dynamic of the goldstone mode is ruled by $-  \frac{a_2^2 k^4}{\Gamma}$, which is very slow. This means that an excitation with a frequency close to the lasing one is damped with a rate very small, thus remains a long time in the system.

To tackle it quantitatively, we take a system with a length $N_y= 100$, comparable to the one of \cite{secli2019theory}. Indeed, we need to make the spacing $\Delta k$ between 2 modes (due to the PBC) small enough to get modes in the flat region. 

\begin{figure}[H]
    \centering
    \includegraphics[scale=0.7]{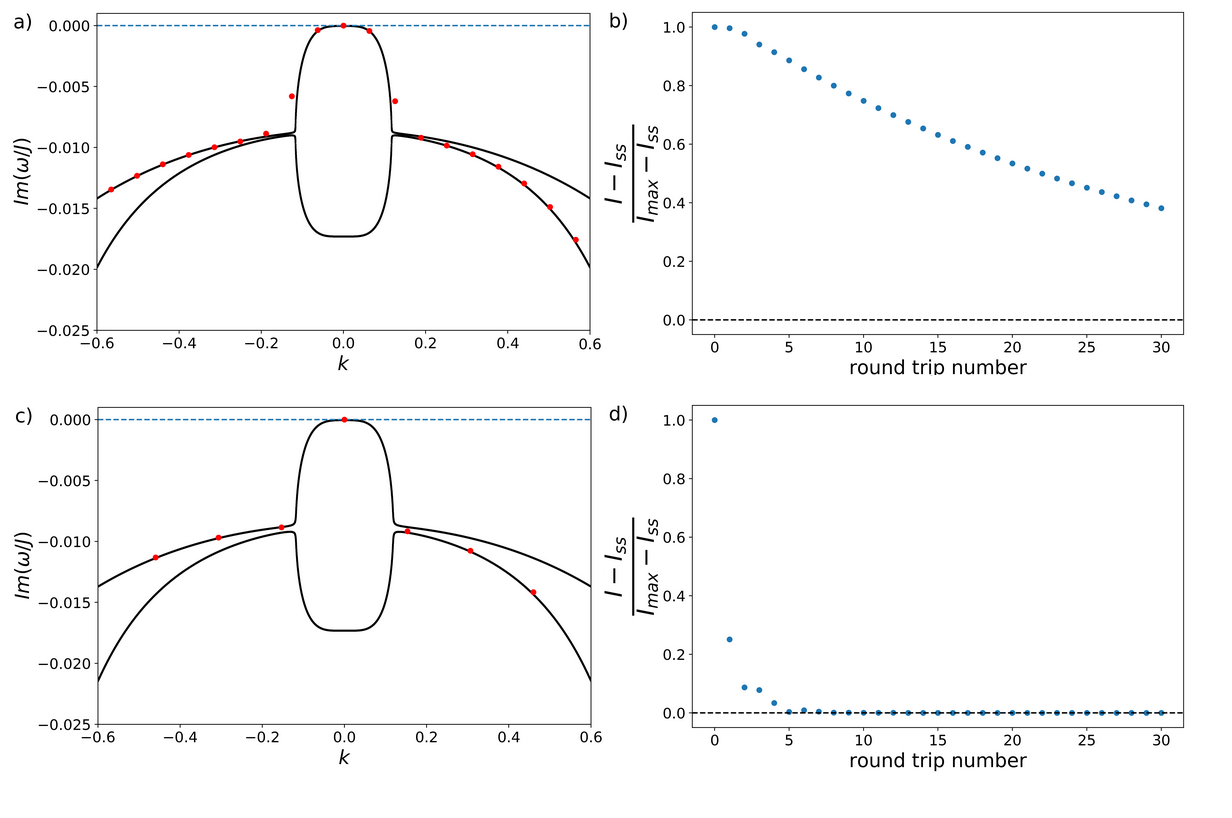}
    \caption{(a),(c) Comparison of the imaginary part of the elementary excitations spectrum between a simulation (red dots) and the 2D model (black line). 
    (b),(d) Evolution of the normalized maximal intensity of the perturbation as a function of the number of round trips in a cylindrical geometry. Parameters: $N_y=100$ in (a),(b), $N_y=41$ in (c),(d).
    }
    \label{fig4.1}
\end{figure}

The procedure is the following:  we let the system evolve until it reaches the steady-state and any perturbation is damped. Then we introduce a little perturbation at a wavevector close to the lasing one and let the system evolve. At each time step after the perturbation, we calculate the space Fourier transform of the system, which shows the time evolution of the intensity of each wavevector of the perturbation. For a given wavevector, we assume this evolution to be an exponential  decay  (which  is  consistent  with  simulations),  and  we  fit  it  to  get  the  slope. This  last  quantity actually corresponds to the imaginary value of the elementary excitations spectrum at a given  wavevector, which allows us to do a direct comparison. The result of this simulation is showed in Fig. \ref{fig4.1} (a) where we see that the agreement is nearly perfect with the 2D model. There are 2 excitations in the flat Goldstone part of this perturbation, which are responsible for the very slow decay of the perturbation whose maximal value decay (normalized to its initial value) is displayed in Fig. \ref{fig4.1} (b) as a function of the round trip number ($\gamma T_{RT} \approx 8 $). 

For completeness, the same operation has been conducted with a smaller lattice, such that there is no excitation in the flat part, as shown in Fig. \ref{fig4.1} (c). In this case, the damping of the perturbation is far quicker, as shown in Fig. \ref{fig4.1} (d).

Finally, it has to be noted that this phenomenon becomes worse and worse when the pump intensity gets higher. Indeed, we saw that the small $|k|$ behaviour of the Goldstone mode is  $- \frac{a_2^2 k^4}{\Gamma}$ where $\Gamma = \gamma(1-\frac{P_{th}}{P})$, so increases with $P$ until saturating at $\Gamma= \gamma$. This way, a system initially quick at a low pump intensity can become very slow if the flat zone of the Goldstone mode extends to an available quantized excitation under the influence of a higher pump intensity.

\section{Reservoir speed tuning}\label{chap4sec2}

To keep on investigating the area of existence of the ultraslow relaxation time, we want to explore what happens when a reservoir is added to the basic case of reference. In this case, we can get an insight easily from the 1D effective model: in Fig. \ref{fig4.3} (a), we see that reducing the speed of the reservoir does not change the small $k$ shape of the Goldstone mode. Thus we expect the ultraslow relaxation time to remain in presence of a slow reservoir.

\begin{figure}[H]
    \centering
    \includegraphics[scale=0.75]{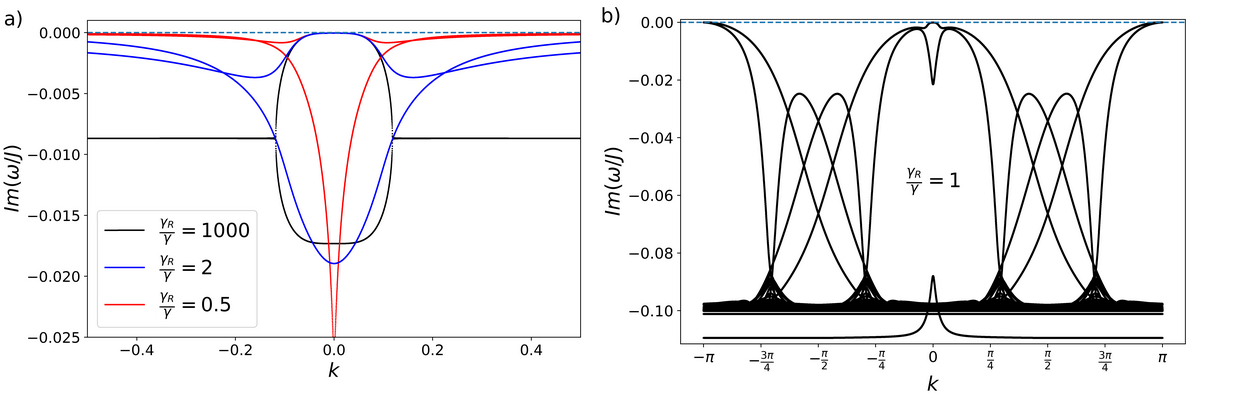}
   \caption{Imaginary part of the elementary excitations spectrum around the lasing edge mode (a) for decreasing reservoir speed ($\frac{\gamma_R}{\gamma} =1000$ $\Leftrightarrow$ adiabatic approximation) computed with the 1D effective model, (b) for  $\frac{\gamma_R}{\gamma} =1$ computed with the 2D model.
   }
    \label{fig4.2}
\end{figure}

At lower reservoir speed, we start to guess in this figure the appearance of another phenomenon. Indeed, as $\gamma_R$ decreases, the branches of modes outside the splitting zone are getting closer and closer from the instability threshold, and increases with $|k|$. Actually, in the real system, this enhancement of the side branches compete with the shape of the gain, and we can see in Fig. \ref{fig4.2} (b) that at large $|k|$ the branches decay. But still, at low $|k|$, this phenomena makes the side branches approaching the instability threshold and  stick to it , extending in a dramatic way the region where perturbations are damped very slowly, and compromising the stability of the system.

Further, this phenomenon also has consequences on the chiral edge modes that propagates in the reverse direction (in blue in Fig. \ref{fig3.1}). Indeed, we just saw that the slow reservoir enhance large $|k|$ excitations. This enhancement is in competition with the shape of the gain, which finishes by winning at very large $|k|$ ($|k| > 1$). But at $|k| \approx \pi$, there are the particle and antiparticle excitations corresponding to the reverse chirality modes. As these modes live on the edge, they have a big overlap with the pump in this region and thus a big gain. This makes that the branches corresponding to these excitations can benefit from the enhancement at large $|k|$ without being suppressed by the shape of the gain, as we can see in Fig. \ref{fig4.2} (b). The consequence of this feature is that when turning on the pump and letting evolve freely, the system will not be able to stabilize in a mode in any of the 2 gaps in a computable time scale, since for the modes of a given gap there is no damping able to suppress the modes in the other gap.

Finally, two ways exist to stabilize the system. The first one consists in pumping at a way higher level. Indeed, in the adiabatic approximation, increasing the pump means increasing $\Gamma$, thus lowering the position of the branches at large $|k|$. This way, it will delay the instabilities to appear when lowering $\gamma_R$, as we can see in Fig. \ref{fig4.3} (b), where for the same value of $\gamma_R$ as in (a) but higher pump intensity the branches are lower. However, this is done at the cost of the enhancement of the ultraslow relaxation rate. 

\begin{figure}[H]
    \centering
    \includegraphics[scale=0.65]{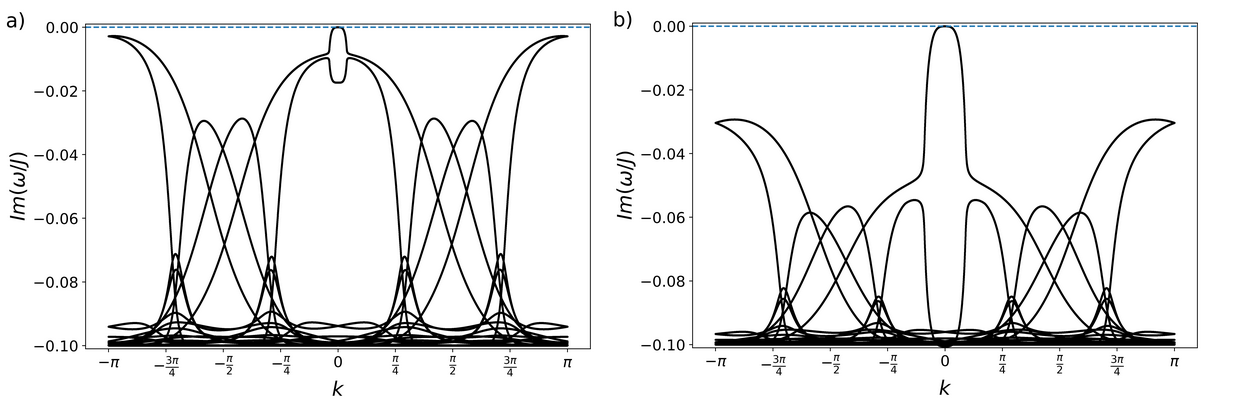}
   \caption{Imaginary part of the elementary excitations spectrum around the lasing edge mode for  $\frac{\gamma_R}{\gamma} =25$ and (a) $P= 1.25\gamma$, (b) $P= 2.25\gamma$.}
    \label{fig4.3}
\end{figure}

The second way to stabilize the lasing in a given mode is starting from a seed, or experimentally putting a little resonant probe. This procedure works, but let the system still fragile in front of any perturbation.

\section{Conclusion}

In this chapter, we applied the models introduced in the previous chapter to explain simply the ultraslow relaxation time of \cite{secli2019theory} in term of slow $k^4$ dynamic of the Goldstone mode around $|k|=0$. We showed that these models give quantitative results and thus allow to predict the adapted size of the lattice to avoid it. A worsening of this relaxation time has also been predicted as pumping intensity is increased.

In a second time, we investigated the same phenomenon but adding a slow reservoir to the system and showed that for moderate values of the reservoir speed it is still valid. However, when the speed of the reservoir is low, we found an enhancement of the gain of the large $|k|$ modes. This behaviour compromises greatly the stability of the system because it creates a large flat zone just behind the instability threshold. It also enhances the gain of chiral edge modes with reverse direction which also become damped very slowly. Finally, we showed that bypassing these problems can be done at the expense of a high pump intensity. It as to be stressed that this behaviour is not model dependent, so should be encountered in any topological laser with chiral symmetry.

Finally, one could ask why we did not investigate the presence of the ultra-slow relaxation time with a non-linear term. The reason is simple: in (\ref{eq:3.7}), if $\mu_{tot} \neq 0$, the limited development around $k=0$ is proportional to $k^2$, namely much quicker than before. This means that the flat zone of the Goldstone mode is narrower, thus removing the ultra-slow relaxation at all the reasonably big lattice sizes. We thus let the analysis of the influence of non-linearities to the next chapter.

\chapter{Influence of nonlinearities}\label{chap5}

In this chapter we want to study the influence of the addition of non-linear terms to our model. A first natural step will be to analyse the role of the sign of the non-linearity, as it is well known that only repulsive (positive interaction strength) interactions preserve a condensate. Then we
will characterize the instability found in \cite{longhi2018presence}. Finally we will unveil some new instabilities which could be found in topological lasers.

\section{Sign of the interaction strength}

From (\ref{eq:3.7}), a limited development at small $|k|$ gives that $Im(\omega_{+}) \approx -\frac{\mu_{tot} a_2 k^2}{\Gamma}$ when $\mu_{tot} \neq 0$. It clearly shows that if $sign(\mu_{tot}) \neq sign(a_2)$ the Goldstone mode is above the instability threshold, making the system unstable.

\begin{figure}[H]   
\begin{minipage}[t]{0.5\textwidth}
It is then interesting to look at the second derivative of the dispersion of the edge modes to see the stability regions for a given sign of the non-linearities. In Fig. \ref{fig5.1}, we see the second derivative of the dispersion of the bottom gap (the one of the top gap is the same with reversed sign) plotted along with the lasing threshold curve with pump intensity $P^{ad}=1.53\gamma$. Up to this pump intensity value, the region that has enough gain to lase has a fully positive mass (the small negative region on the right cannot be reached in practice because is too close to the border of the gap).
\end{minipage}
\begin{minipage}[t]{0.5\textwidth}
    \centering
    \raisebox{\dimexpr-\height+\ht\strutbox\relax}{\includegraphics[scale=0.35]{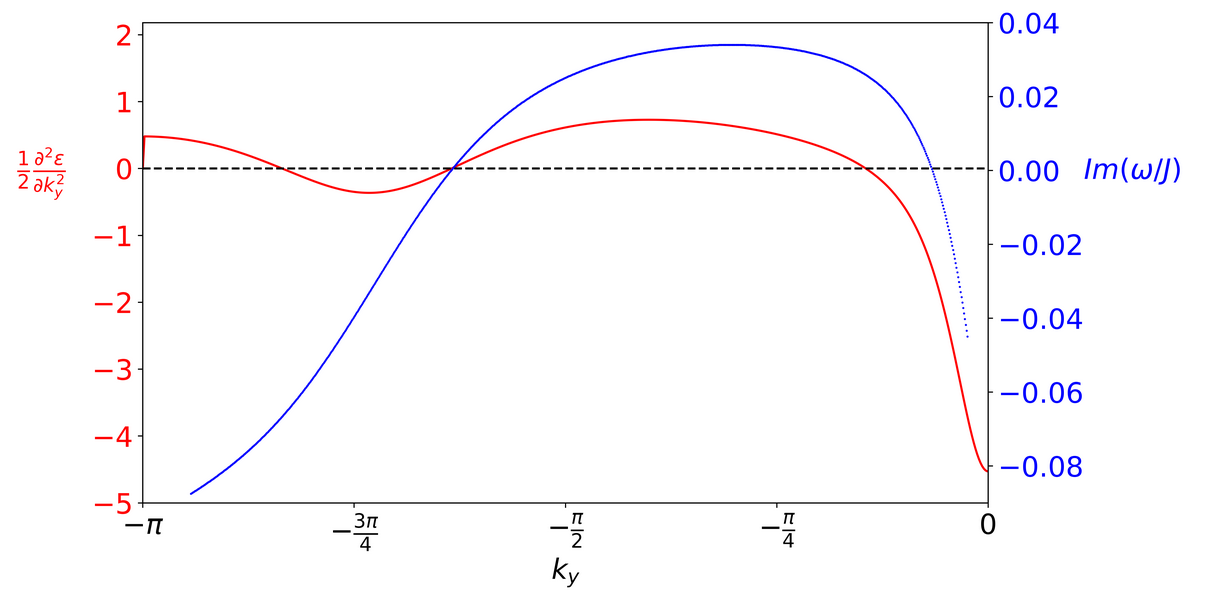}}
     \caption{Second derivative (red) of the dispersion of the bottom gap versus lasing threshold profile (blue) with $P^{eff}= 1.53\gamma$ (and $\gamma=0.1J$) as a function of the wavevector $k_y$.
    }\label{fig5.1}
\end{minipage} 

\end{figure}

This means that if $sign(\mu_{tot}) > 0$, all the modes available are stable in this gap, but unstable in the other one. This way, the chiral symmetry of the lattice is effectively broken by the presence of a non-linearity.

\section{Instabilities in \cite{longhi2018presence}}

In this section, we want to answer the natural question of the origin of the instability unveiled in the article explained in Chapter \ref{chap2} section \ref{chap2sec2}. 

\begin{figure}[H]   
\begin{minipage}[t]{0.5\textwidth}
We can see in the insets that the cases where
the 

other gap modes are involved are just a minority. In most cases, the excitations involved in the perturbation are in the same gap as the dominant mode, which is still the same mode as the one lasing when the reservoir is fast. However in presence of a non-linearity, the Goldstone mode is supposed to be very narrow. It means that this kind of instability is different from what we have seen so far and needs to be studied more in detail. To this end, we compute the elementary excitations spectrum with these parameters. We use the fact that there is a small stable window after a change of parameters to compute it. The result is showed in Fig. \ref{fig5.3}: in (a), we can see that the Higgs mode is now strongly hybridized with reservoir modes and gets its side branches overcoming the instability threshold. This confirms that the modes at the origin of the instability are situated in the same gap as the lasing mode, in coherence with \cite{longhi2018presence}. Also, we see that the branches of the reverse chirality modes are still quite enhanced and close to the threshold, which explains that these modes can be involved in some case in the presence of disorder. In (b), we can see that the 1D effective model gives results that are qualitatively in agreement with the 2D model, which makes this former a good tool to get rough estimates of the values of $\gamma_R$ at which the system starts to be unstable. Finally, the proximity between this model and the one commonly used for non-equilibrium condensates gives us the physical explanation of this phenomena. As observed in \cite{wouters2007excitations}, this instability is due to the repulsive interactions between the lasing mode particles and reservoir ones: "a local depletion of the reservoir density $N_n$ creates a potential well which attracts the lasing mode particles. This in turn leads to a further drop of the local reservoir density by the hole-burning effect, until a spatially modulated steady-state is eventually reached."
\end{minipage}.
\begin{minipage}[t]{0.5\textwidth}
    \centering
    \raisebox{\dimexpr-\height+\ht\strutbox\relax}{\includegraphics[scale=0.95]{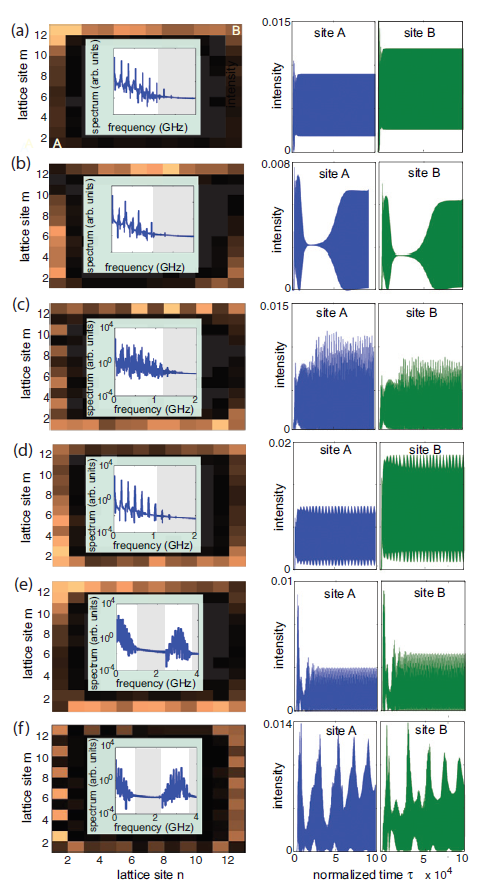}}
     \caption{Same as Fig. \ref{fig0.3}, but with disorder of resonance frequencies $\omega_{ref}$ of each site. The panels, from (a) to (f), show different runs corresponding to different realizations of disorder, displaying distinct dynamical behavior. The insets in the left panels show the power spectrum of the temporal laser intensity at site A. The frequency is normalized to the one of the stable lasing mode when the reservoir is fast. The gaps are indicated by the white background. In (e) and (f), the spectral peaks at higher frequencies (at around $\sim$ 3GHz) are ascribed to the beating of counterpropagating chiral edge modes. In 30 runs, steady-state oscillation after initial transient is observed in 4 runs (i.e., less than 20\% times). The most probable outcome is the one in panel (a) (stable periodic limit cycle). Adapted from \cite{longhi2018presence}.
    }\label{fig5.2}
\end{minipage} 
\end{figure}

The first idea that comes to mind in the light of what we have done so far is that the origin of this instability could be linked with the reverse chirality modes of Chapter \ref{chap4} section \ref{chap4sec2}. To get more information about the results of \cite{longhi2018presence}, in Fig. \ref{fig5.2} we displayed further simulations presented in the article, which show the different results obtained when running the system with a low reservoir speed in presence of small disorder. 

\begin{figure}[H]
    \centering
    \includegraphics[scale=0.65]{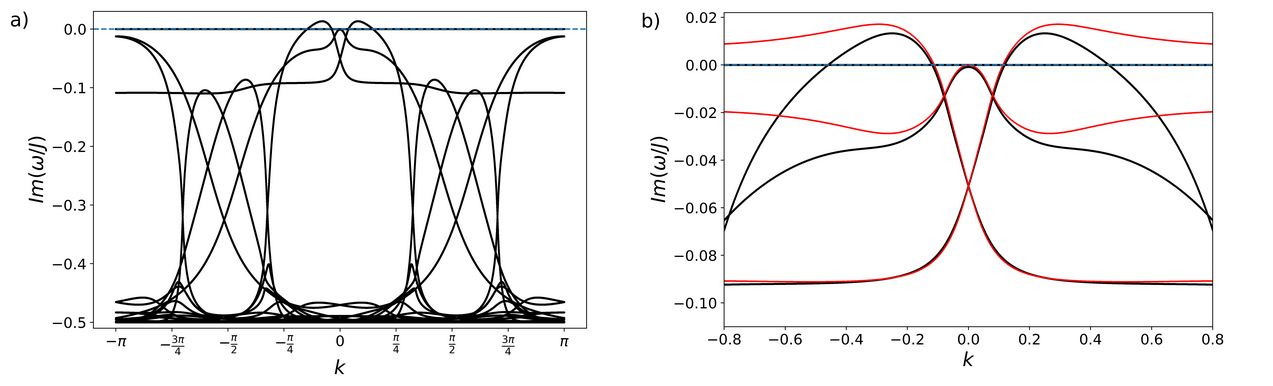}
    \caption{Imaginary part of the elementary excitations spectrum with same parameters as Fig. \ref{fig0.3} (without output coupler). (a) panel: full spectrum from the 2D model. (b) panel: zoom on the center of the spectrum to compare the results from the 2D model (black lines) and the ones from the 1D effective model (red lines, with full edge modes dispersion taken into account).
    }
    \label{fig5.3}
\end{figure}

\section{Reverse chirality instability}

In the previous section, we explained the instability due to a slow reservoir in presence of a non-linear term. However, when playing with the reservoir speed parameter, we found another striking phenomena.

\begin{figure}[H]   
\begin{minipage}[t]{0.5\textwidth}
Indeed, at intermediate values of $\gamma_R$ (all other parameters kept constant), we find an unstable regime in which mainly modes with opposite chirality are involved. We show  in Fig. \ref{fig5.4} the elementary excitations spectrum corresponding to this case, where we clearly see that the branches corresponding to modes with opposite chirality are above the instability threshold.
\end{minipage}
\begin{minipage}[t]{0.5\textwidth}
    \centering
    \raisebox{\dimexpr-\height+\ht\strutbox\relax}{\includegraphics[scale=0.67]{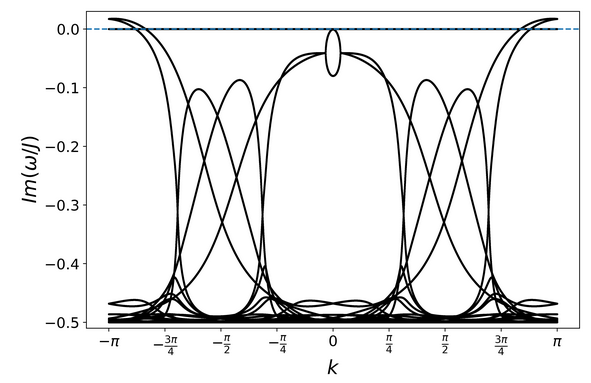}}

\end{minipage} 
     \caption{Imaginary part of the elementary excitations spectrum with same parameters as Fig. \ref{fig0.3} except $\frac{\gamma_R}{\gamma}= 2$ (without output coupler).
    }\label{fig5.4}
\end{figure}

\begin{figure}[H]   
\begin{minipage}[t]{0.5\textwidth}
To study further this effect, we come back to a simpler case, taking the parameters of reference and just adding a non-linear term under the form of $g > 0$. In our simulations, when the system is let free to evolve to a SS, we find that above $g \approx 0.035$ ($\iff \mu_{tot} = 0.2 J \approx 0.13 \Delta \omega_{gap}$), the system is not stable anymore. Just behind the threshold of this instability, we plot in Fig. \ref{fig5.5} the imaginary part of the elementary excitations spectrum.
\end{minipage}
\begin{minipage}[t]{0.5\textwidth}
    \centering
    \raisebox{\dimexpr-\height+\ht\strutbox\relax}{\includegraphics[scale=0.65]{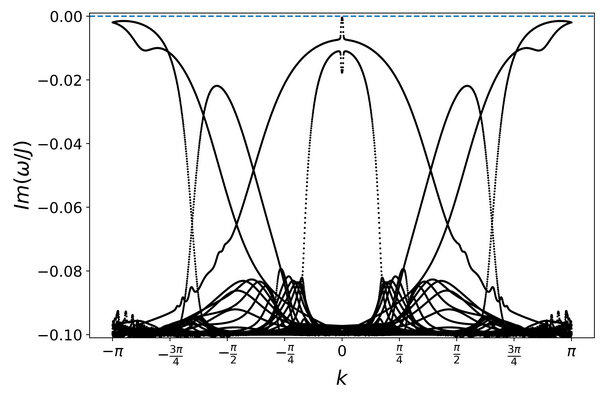}}
\end{minipage} 

   \caption{Imaginary part of the elementary excitations spectrum for $g=0.035$ computed with the 2D model. Parameters of reference.}
    \label{fig5.5}   
\end{figure} 

\vspace{0.5cm}
We see on this figure that the particle and antiparticle modes corresponding to the mode of opposite chirality on the same edge are very close from the instability threshold.

For this new type of instability, we succeeded in finding a qualitative physical explanation. Indeed, as the strength of the non-linearity is raised, the modes of the bare HH Hamiltonian undergo changes in their spectral and spatial shapes, and so do the elementary excitations. 

\begin{figure}[H]   
\begin{minipage}[t]{0.45\textwidth}
In particular, if we look at the spatial localization of the Goldstone modes and of the reverse chirality modes, as shown in Fig. \ref{fig5.67}, we can see that their localization (we focus on the Goldstone and reverse chirality mode which are the most localized on the edge) on the edge of the lattice change compared to the bare HH model (where this localization is the same), and in a different way. The consequence of this is that the overlap between the reverse chirality modes and the pump intensity is bigger than for the Goldstone modes, thus these former receive more gain than these latter. 
\end{minipage}
\begin{minipage}[t]{0.55\textwidth}
    \centering
    \raisebox{\dimexpr-\height+\ht\strutbox\relax}{\includegraphics[scale=0.7]{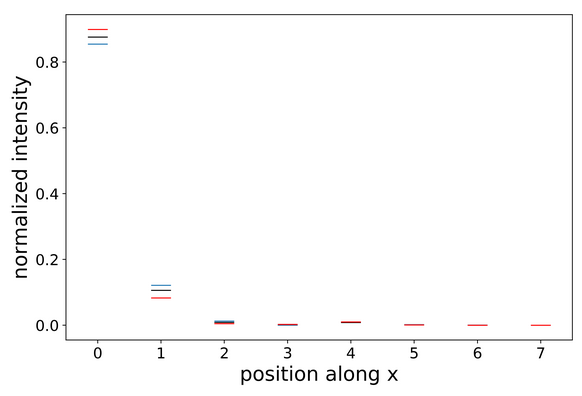}}
    \caption{Normalized spatial localization of a Goldstone (blue bars) and a reverse chirality mode(red bars) mode  for $g=0.035$ and both type of modes (black bars) for $g=0.0$ along the $x$-axis. The two modes are chosen such that they are the one of their branch with the maximal overlap with the edge ($x=0$).} 
    \label{fig5.67} 
\end{minipage} 
\end{figure}

It is thus normal that the reverse chirality excitations branches are enhanced compared to the branches aside the Goldstone mode, and possibly cross the instability threshold at a certain point. Finally, note that in this case the 1D effective model is by construction (with the bare HH Hamiltonian eigenvectors) inefficient.

\section{High pump intensity instabilities}

In this last section, we unveil a new type of instability linked with the presence of a non-linear term. Indeed, another way to increase the blueshift due to the non-linear term is to increase the pump intensity (as $\mu_{tot} = g G_0 |\psi|^2$, where $|\psi|^2$ increases linearly with the pump intensity according to \ref{chap3sec7}). 

\begin{figure}[H]
    \centering
    \includegraphics[scale=0.75]{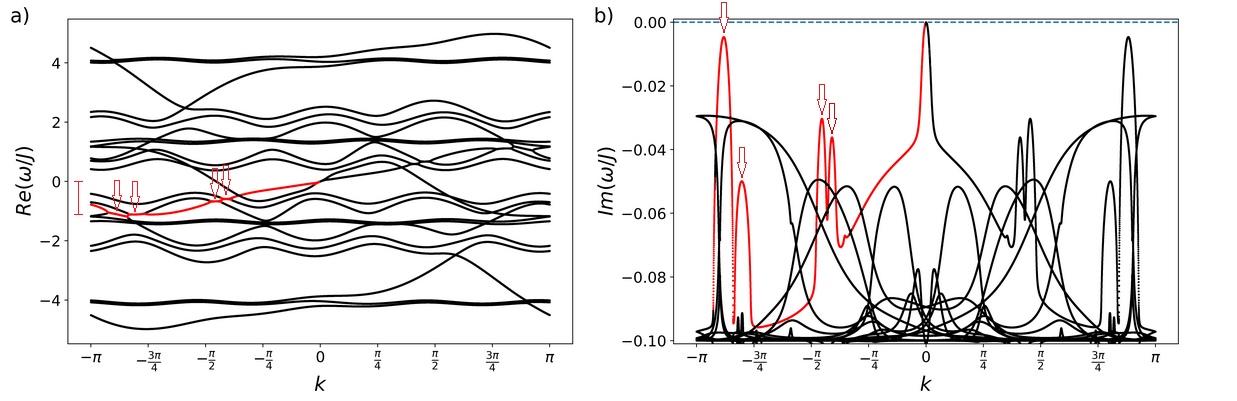}
   \caption{Real (a) and imaginary (b) part of the elementary excitations spectrum as a function of $k=k_y - k_{y,0}$, for $g=0.01$ and $P^{ad}= 2.15\gamma$ computed with the 2D model. The red line indicates the edge modes branch. The arrows indicate the interacting sections of the branch. The red vertical line in (a) indicates the position of the most interacting section relatively to the lasing edge mode. }
    \label{fig5.7}  
\end{figure}

Here again, above a certain value of the pump intensity, the system is not able anymore to reach a SS. In Fig. \ref{fig5.7}, we plotted the imaginary part of the elementary excitations spectrum just behind the instability threshold ($P^{ad}= 2.15\gamma \iff \mu_{tot} = 0.6 J \approx 0.4 \Delta \omega_{gap}$) with the 2D model. 
We see in (b) that the branch of the edge modes presents bumps (red arrows) corresponding in (a) to the section of interaction between edge modes and other modes. In particular, the bump with the highest value in (b) corresponds to the interaction between edge modes and modes inside the trivial gap (the ones in blue in the central gap of Fig. \ref{fig2.1}), which are also localized on the edge. This way, the overlap of the resulting modes is also important with the edge, making them getting a lot of gain, which explains they are the first one able to destabilize the system.

Finally, the origin of this kind of instabilities is the strong change in the modes structure of the system induced by the non-linearity. The interaction of trivial modes and edge modes is reinforced as the pump intensity increases and leads to the appearance of modes able to destabilize the system. As this phenomenon involves trivial modes, it is thus hard to explain and to predict (the 1D effective model is obviously totally inefficient in this case), and also very model dependent.

\section{Conclusion}

In this chapter, we studied the influence of non-linearities on the system. A first major physical consequence is that non-linearities effectively break the chiral symmetry of the system, by making dynamically unstable one of the two chiralities. Then, the case of \cite{longhi2018presence} was studied, leading to the definition of a new kind of instability arising from the strong interaction of reservoir excitations with the one of the lasing edge mode. The 1D effective model proofed to be efficient to predict it.

We then reduced the influence of the reservoir and discovered two new instabilities. One is due to the enhancement of reverse chiral edge modes by the spreading of the lasing edge mode in the transverse direction of the lattice at high non-linear strength. The other is due to interactions at high pump intensity between edge modes and trivial modes.  

Overall, this study shows that the dynamical stability of topological lasers could be severely compromised in non-linear materials. Indeed, we found instabilities in any type of scenario, making the stable range of parameters narrow.

\chapter*{Conclusion}
\addcontentsline{toc}{chapter}{Conclusion}

The work exposed in this report is a contribution to the understanding of the dynamical stability of 2D topological lasers. It is up to date the most complete description of the diverse phenomena at play in this kind of devices. Indeed, we started from the existing literature, and explained in chapter \ref{chap4} the ultra-slow relaxation time in \cite{secli2019theory} in term of the flatness of the Goldstone mode around its center. This phenomenon has been predicted to exist for big enough lattices without non-linearity. The presence of a reservoir keeps this phenomenon unchanged at intermediate career relaxation time, while the system becomes nearly unstable at low reservoir speeds because of the enhancement of the gain of the reverse chirality modes which become slowly damped as well. 

We then studied the influence of non-linearities on the system, unveiling several major features. First, the chiral symmetry is effectively broken by non-linearites, because for a given sign of the non-linearities, only one gap has the appropriate mass sign to allows for dynamical stability. Second, the instability found in \cite{longhi2018presence} has been explained, originating from the strong interplay between a slow reservoir and non-linearities. Third, strong non-linearities were shown to create other instabilites, either by enhancing the reverse chirality modes by a mechanism of spreading of the lasing mode, either by the enhancing of coupled bulk modes. This way, we establish the stable phase to be reduced to systems with a fast reservoir and small non-linearities. 

Finally, the 1D effective model for a lasing topological edge mode we derived is also a major advance since it allows for a simple, quantitative and universal (it depends only on the pump threshold and the dispersion coefficients of a given model) understanding of most of the phenomena depicted. 

\textbf{Outlooks:}

The previous results represent the foundation of the study of the stability of 2D topological lasers, and the methods we developed can now be applied to study further phenomena. For example, in \cite{secli2019theory}, the question of topological lasing under a finite size pumping spot has been explored, showing notably the existence of a convective instability threshold before the absolute instability one. Further properties in this geometry could be explored easily by mean of their elementary excitations. 

Another interesting question could also be the investigation of the spatial coherence of this kind of lasers, in order to predict the spatial size of these type of non-equilibrium condensates in presence of spatial and temporal fluctuations. In this case, the1D effective model we introduced and its similarity with the one depicting non-equilibrium condensates in planar cavities could be used to revisit previous studies of these latter \cite{chiocchetta2013non, gladilin2014spatial, squizzato2018kardar}.

\textbf{Acknowledgements:}

My acknowledgments go to all the BEC center members, who all participate in the welcoming atmosphere of this lab, and in particular to my advisor Iacopo Carusotto whose eternal optimism and joviality, as well as his big physical knowledge and intuitions helped my so much during these last 9 months. I also thank especially Ivan Amelio for his precious help all along this project and Matteo Secli for his codes.

\renewcommand\bibname{References}
%\begin{changemargin}{-2cm}{-2cm}
%\begin{multicols}{2}
%\let\clearpage\relax
%\fontsize{10}{10}\selectfont{
\bibliographystyle{unsrt}
\bibliography{ref}   %}
%\end{multicols}
%\end{changemargin}
%\addcontentsline{toc}{chapter}{References}

% apendices

%\input{chapters/appendix_A.tex}
%\input{chapters/appendix_B.tex}
\begin{appendices}

\renewcommand{\thechapter}{A}
\chapter{Appendix}\label{appendixA}

\begin{changemargin}{-2 cm}{-2cm} 
\begin{multicols}{2}
\fontsize{10}{10}\selectfont{

\subsubsection{Equations of evolution}

Full 2D equations:

\begin{equation}\label{eq:A1}
\begin{split}
      i\frac{\partial \psi_{m,n}(t)}{\partial t} = &-J\Big[\psi_{m+1,n} + \psi_{m-1,n}  + e^{-2\pi i \theta m}\psi_{m,n+1} \\
      & + e^{+2\pi i \theta m}\psi_{m,n-1}\Big]  + \Big[ g|\psi_{m,n}|^2  + g_R N_{m,n} \\
      & + i(R N_{m,n} - \gamma) \Big] \psi_{m,n}     
\end{split}
\end{equation}

where the reservoir density $N_{m,n}$ is determined by the rate equation,

\begin{equation}\label{eq:A2}
      \frac{\partial N_{m,n}}{\partial t} = P \delta_{m,0} - (\gamma_R + R|\psi_{m,n}|^2)N_{m,n} 
\end{equation}

Formally, we can rewrite (\ref{eq:2.1}) as:

\begin{equation}\label{eq:2.3}
    i\frac{\partial \ket{\psi}}{\partial t} = (H^{HH} + V) \ket{\psi} 
\end{equation}

where $H^{HH}$ corresponds to the Harper-Hofstadter contribution (thus Hermitian) and $V$ is a contribution including nonlinear and non-hermitian terms.

First, due to the ill nature of $V$, it is in general not possible to look for a solution of this problem using the eigenvectors of $H^{HH}$ as a basis. However, we make the approximation that when $V$ is small it is possible to write a solution using these later. Further, we restrict our basis to the eigenvectors of the edge modes of a given gap, as we guess it is sufficient to retrieve the interesting elementary excitations. We thus write generally a solution as:

\begin{equation}\label{eq:2.4}
    \braket{m,n}{\psi} = \sum_{k_y} A(k_y) \braket{m,n}{\lambda^{k_y}} = \sum_{k_y} A(k_y) e^{i k_y n} \phi^{k_y}_m
\end{equation}

where $H^{HH} \ket{\lambda^{k_y}} = \lambda^{k_y} \ket{\lambda^{k_y}}$. The second equality is due to periodicity of the lattice along the $y$-axis (application of Bloch theorem).

Second, as a lasing chiral edge mode is mainly concentrated on the gainy edge of the lattice, we focus on the $\psi^{1D} = \psi_{0,n}(t)$ function:

\begin{equation}\label{eq:2.5}
    \braket{n}{\psi^{1D}} = \psi_{0,n} = \sum_{k_y} A(k_y)\phi^{k_y}_0 e^{ik_y n}
\end{equation}

namely the time evolution of the amplitude on the last site in the edge.

This function can be obtained from the 2D one by summing at a given $k_y$ in Fourier space over:

\begin{equation}
    \rightarrow \sum_m ... \times \phi^{k_y, norm}_m G(k_y)
\end{equation}

where $G(k_y) = |\phi_{0}^{k_y}|^2 $ and $\phi^{k_y, norm}_m = \frac{\phi^{k_y}_m}{\phi^{k_y}_0}$ (since $\phi^{k_y}_m$ is normalized as $\sum_m |\phi^{k_y}_m|^2 = 1$ \ $\forall k_y$), which amounts to project the equation on the last site of the lattice.

\subsubsection{Derivation of the 1D equation}
\begin{itemize}

\item First, we remark that the equation (\ref{eq:A2}) has a non
zero SS only for $m = 0$. Thus we make the approximation that
$N_{m,n} = N_n \delta_{m,0}$ at any time, which lets the equation unchanged for $\psi^{1D}$.

\item Then in Fourier space, the $H^{HH}$ term in (\ref{eq:A1}) gives:
    
    \begin{equation}
      \bra{m,k_y}  H^{HH} \ket{\psi} = A(k_y) \epsilon(k_y) \phi^{k_y}_m
    \end{equation}
    where $\epsilon$ is the edge modes dispersion in a given gap.
    
    Taking into account that the $\phi^{k_y}$ are normalized along the $x$-axis (but not orthogonal in $k_y$) then:
    
    \begin{equation}
        \sum_{m} A(k_y) \epsilon(k_y) |\phi^{k_y}_m|^2 = A(k_y) \epsilon(k_y)
    \end{equation}
    
    Thus:
    
    \begin{equation}
        \mel{n}{H^{HH}_{1D}}{\psi^{1D}} = TF^{-1}[\epsilon] \circledast \psi^{1D}_n
    \end{equation}
     where $\circledast$ stands for the convolution product.

\item  in fourier space, the $N\psi$ term reads:

 \begin{equation}
         [N(k_y)] \circledast [A(k_y) \psi^{1D}(k_y)] (k_y')
    \end{equation}
    
    Then the projection is only the multiplication of the precedent expression by $G(k_y') \phi^{k_y, norm}_0 = G(k_y')$.
    
    Thus coming back to real space it gives:
    
    \begin{equation}
        = \Big[TF^{-1}[G] \circledast [N_n \psi^{1D}_n] \Big]_n
    \end{equation}
    
\item for the $g|\psi_{m,n}|^2 \psi_{m,n}$ term, we first note that for $k_y$ close to the gain maximum, $|\phi^{k_y, norm}_{1}|^2 \approx 0.1$ (see Fig. \ref{figA.1}). 
    
Then in Fourier space and over projection along $x$ it reads:
    
\begin{equation}
g \sum_m \Big[ TF[ |\psi_{m,n}|^2]\circledast [\psi_{m,k_{y'}}] \Big]_{k_y} \phi^{k_y,norm}_m G(k_y)
\end{equation}
    
For each $m$, it amounts approximately to multiply four $\phi^{k_y,norm}_m$ functions between each other. So close to the gain maximum, we can approximate that the $m>0$ terms are negligible.
    
Therefore the final result in real space is:
    
\begin{equation}
g TF[G]\circledast [|\psi^{1D}|^2] \psi^{1D}|
\end{equation}

\end{itemize}

\subsubsection{Full 1D effective equations:}

Finally, we can write the 1D equation of evolution of a lasing edge mode $\psi_n$ (dropping the $^{1D}$ label):

\begin{equation}\label{eq:A22}
      \left \{
\begin{array}{l}
\begin{split}
&i\frac{\partial \psi_n}{\partial t} =  TF^{-1}[\epsilon(k_y)] \circledast \psi_n + i\big( R\times TF^{-1}[G(k_y)] \circledast [N_n \psi_n] - \gamma \psi_n \big)\\ 
&+ g TF^{-1}[G(k_y)] \circledast [|\psi_n|^2\psi_n] + g_R TF^{-1}[G(k_y)] \circledast  [N_n \psi_n]
\end{split}\\
\frac{\partial N_{n}}{\partial t} = P - (\gamma_R + R|\psi_n|^2)N_{n}
\end{array}
\right.
\end{equation}

where $\psi_{N_y} = \psi_{0} \ \forall m$, $N_{N_y} = N_{0} \ \forall m$.

\subsubsection{Shape of the gain and lasing threshold:}

We now want to give more insights about the $G(k_y)$ term. First,we see that this function corresponds to the overlap between a mode and the edge. For the driven-dissipative term, it means that the gain is not momentum uniform but rather depends strongly on it.

Second, in Fig. \ref{figA.1}, we see the shape of the $G(k_y)$ function for the edge states along $x=0$ of the bottom gap. An important thing to remark is that the overlap is never equals to 1, due to the penetration of the wavefunction in the bulk. It explains why the lasing threshold can never be equal to $\gamma$. Moreover, it allows to calculate in a simple manner the threshold.

\begin{figure}[H] 
\centering
\includegraphics[scale=0.75]{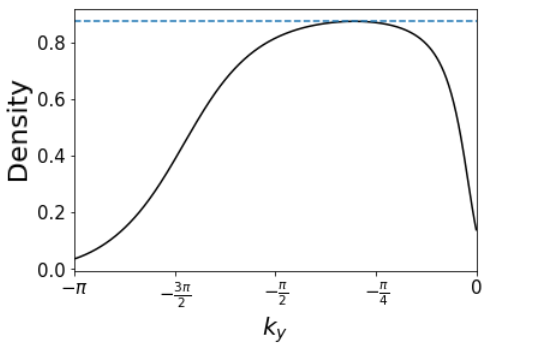}
\caption{Overlap between the edge modes density and the edge ($x=0$). The edge modes are the one of the bottom gap of the bare HH Hamiltonian, and are indexed by the quantum number $k_y$. }
\label{figA.1}   
\end{figure} 

Indeed, let us suppose that we want to reach a density uniform steady-state $\psi_n(t)= \psi^0 e^{-i(\mu t - k_{y,0} n)}$ with a reservoir adiabatically eliminated. We thus have that $G(k_y) TF[N_n \psi_n] = G(k_y) N^0 \psi^0 \delta(k_y - k_{y,0})$ so  $TF^{-1}[G(k_y)] \circledast [N_n \psi_n] = G(k_{y,0}) N^0 \psi_n$. The gain/loss term in (\ref{eq:A2}) then reads $i\Big(\frac{\beta G(k_{y,0}) P}{1+\beta |\psi^0|^2} - \gamma \Big)\psi_n$ which shows clearly that the threshold corresponds to:

\begin{equation}
     P^{ad}_{th}(k_{y,0})= \frac{\gamma}{ G(k_{y,0})}  
\end{equation}

since the intensity is zero at threshold by continuity. As an example, if we calculate the threshold at the maximum value of the overlap (blue dashed line in Fig. \ref{figA.1}), we retrieve the threshold value that we found in Fig. \ref{fig2.2}. We also find again that the system will tend naturally (if not forced with a seed) to lase in modes in the middle of the gap, where the overlap is maximal.

\textbf{Cooking recipe:}

When $\frac{\gamma}{J}$ becomes comparable to 1, the approximation we did to project on the bare HH eigenvectors breaks down. However, as we saw in the previous paragraph the zero order of the $G$ function gives the lasing threshold, and we have an algorithm to calculate the lasing threshold of the full 2D model. We thus can infer the zero order of the $G$ function at high $\frac{\gamma}{J}$ from this new threshold. We then suppose that the shape of the $G$ function remains the same (at least at intermediate $k$, i.e. $|k| \approx 0.75$), so we set:

\begin{equation}
  G^{cook}(k) = \frac{\gamma}{P^{algo}_{th}(k_{y,0})}\frac{G(k)}{G(k_{y,0})}  
\end{equation}

where $P^{algo}_{th}$ is the threshold calculated with the algorithm of chapter \ref{chap1} section \ref{chap1sec3}.

\textbf{Minimal model: 1D generalized Gross-Pitaevskii Equation}

The model presented in the last part can be further simplified. Indeed, as shown in Fig. \ref{fig3.1}, the relevant information on the stability of the system, namely the weakly negative excitations, is mostly limited to the branch of the Goldstone modes. This means that the large $k_y$ behaviour of the edge mode dispersion $\epsilon$ and the overlap function $G$ is superfluous and thus can be neglected. 

Keeping only the 0 order of the overlap function at the lasing point $G(k_{y,0}) \equiv G_0$, the equations of evolution of a lasing edge mode $\psi^{1D}(t) = \psi_n e^{i(\mu t - k_{y,0}n)}$ take the form of generalized Gross-Pitaevskii equations \cite{carusotto2013quantum}:

\begin{equation}\label{eq:2.11}
      \left \{
\begin{array}{l}
\begin{split}
    i\frac{\partial \psi_n}{\partial t} = &\Big[ TF^{-1}[\epsilon(k_y)] \circledast \psi_n + g G_0 |\psi_n|^2 
    + g_R G_0 N_n \\
    &+ i\Big(R G_0 N_n - \gamma\Big)\Big]\psi_n
\end{split}\\
    \frac{\partial N_n}{\partial t} =
    P - (\gamma_R +R |\psi_n|^2) N_n
\end{array}
\right.
\end{equation}

\textit{}
}
\end{multicols}
\end{changemargin}

\end{appendices}
\begin{appendices}

\renewcommand{\thechapter}{B}
\chapter{Appendix}\label{appendixB}

\begin{changemargin}{-2 cm}{-2cm} 
\begin{multicols}{2}
\fontsize{10}{10}\selectfont{

\textbf{Lasing mode tuning}

In this section, we show the comparison between the models when the lasing points are on the edge of the gain distribution.
\begin{figure}[H]   
    \centering
    \raisebox{\dimexpr-\height+\ht\strutbox\relax}{\includegraphics[scale=0.45]{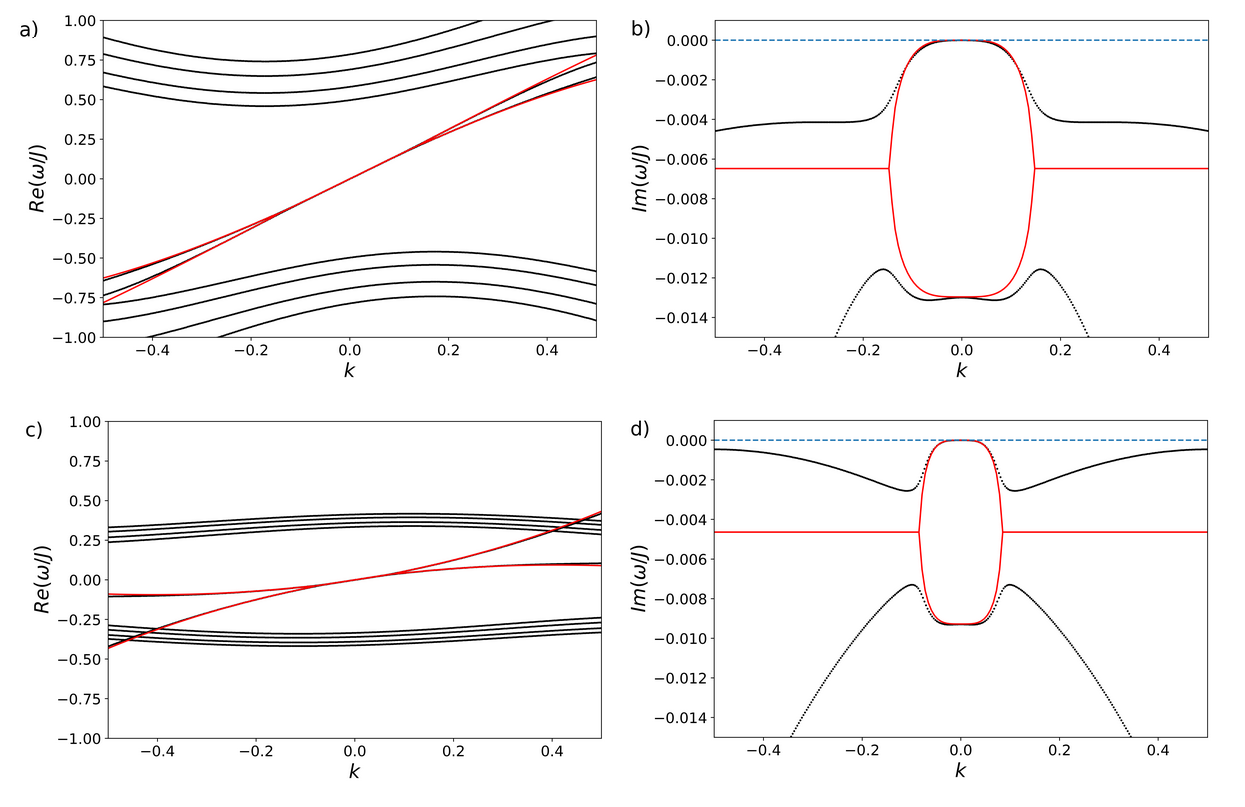}}
        \caption{Real (a),(c) and imaginary (b),(d) part of the elementary excitations spectrum for the 2D model (black dots) and the gGPE model. (a),(b) panels: the lasing mode is \{$k_{y,0}=  -0.6130$, $\mu = -1.5104 J$ \}. (c),(d) panels: the lasing mode is \{$k_{y,0}=  -1.4559$, $\mu = -2.4022 J$ \}
    }
\end{figure}   

We see that the agreement between the 2D model and the gGPE model is good concerning the Goldstone and Higgs modes. However the information contained in the side branches is missed, though it has an importance in this case. This is why it is not relevant to use the gGPE model oustide the maximum of the gain.

\textbf{Reservoir speed tuning}

This time, we compare the different models taking a very low value of the ratio $\frac{\gamma_R}{\gamma}$. In this case, it can be useful to keep the full dispersion of the edge modes (as this will be in Chapter \ref{chap4} section \ref{chap4sec2}). Again the result is very satisfying concerning the Goldstone mode.

\begin{figure}[H]
\centering
    \raisebox{\dimexpr-\height+\ht\strutbox\relax}{\includegraphics[scale=0.45]{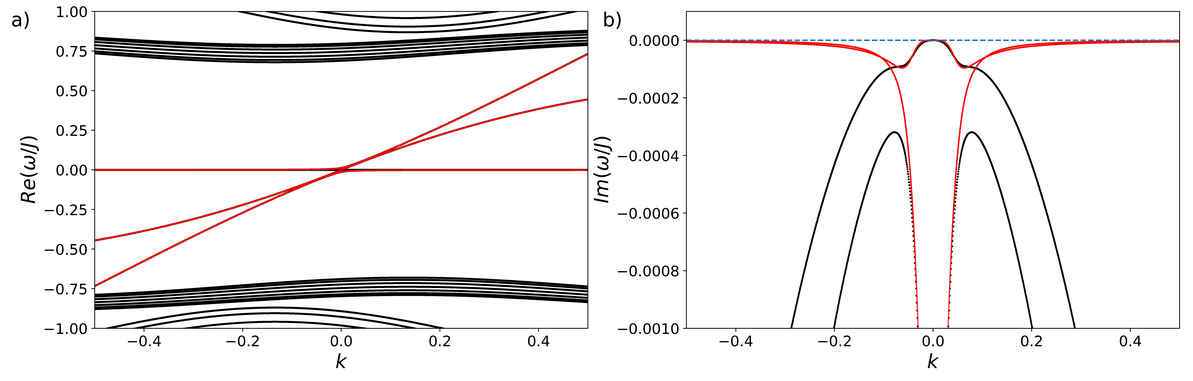}}
    \caption{Real (a) and imaginary (b) part of the elementary excitations spectrum for the 2D model (black dots) and the gGPE model (red lines) for $\frac{\gamma_R}{\gamma}=0.1$
    }
\end{figure}

\textbf{Non-linear term}

This time, we compare the different models by adding non-linearities, taking $g=0.01$, which corresponds to a blueshift of $\mu_L= 0.086$ ($\approx 5 \%$ of the gap). The gGPE model captures perfectly the Goldstone mode. Note that soon above this value of $g$, some problems are appearing, as studied in Chapter \ref{chap5}.
\begin{figure}[H]  
\centering
    \raisebox{\dimexpr-\height+\ht\strutbox\relax}{\includegraphics[scale=0.45]{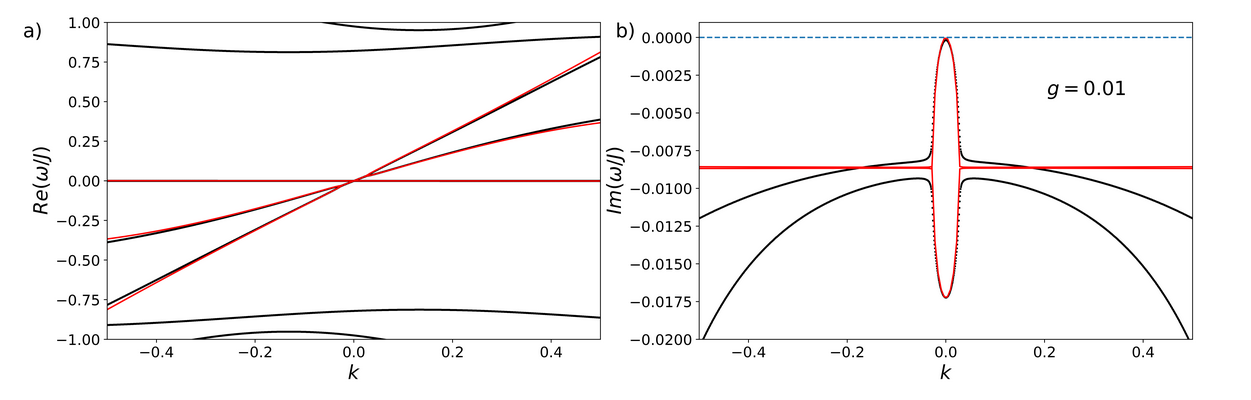}}
     \caption{Real (a) and imaginary (b) part of the elementary excitations spectrum for the 2D model (black dots) and the gGPE model (red lines) for $g=0.01$.
    }
\end{figure}

\textbf{Cooking recipe at high pump intensity}

This time, we compare the different models increasing the value of the pump intensity (via $\gamma$). As we see, for $\gamma=1.5J$, the agreement between the 2D model and the gGPE one (using the cooking recipe) is still very good at the level of the Goldstone and Higgs modes. Rising further up to $\gamma=5.25J$ the agreement remains still good, allowing for quantitative predictions.
\begin{figure}[H] 
    \centering
    \raisebox{\dimexpr-\height+\ht\strutbox\relax}{\includegraphics[scale=0.45]{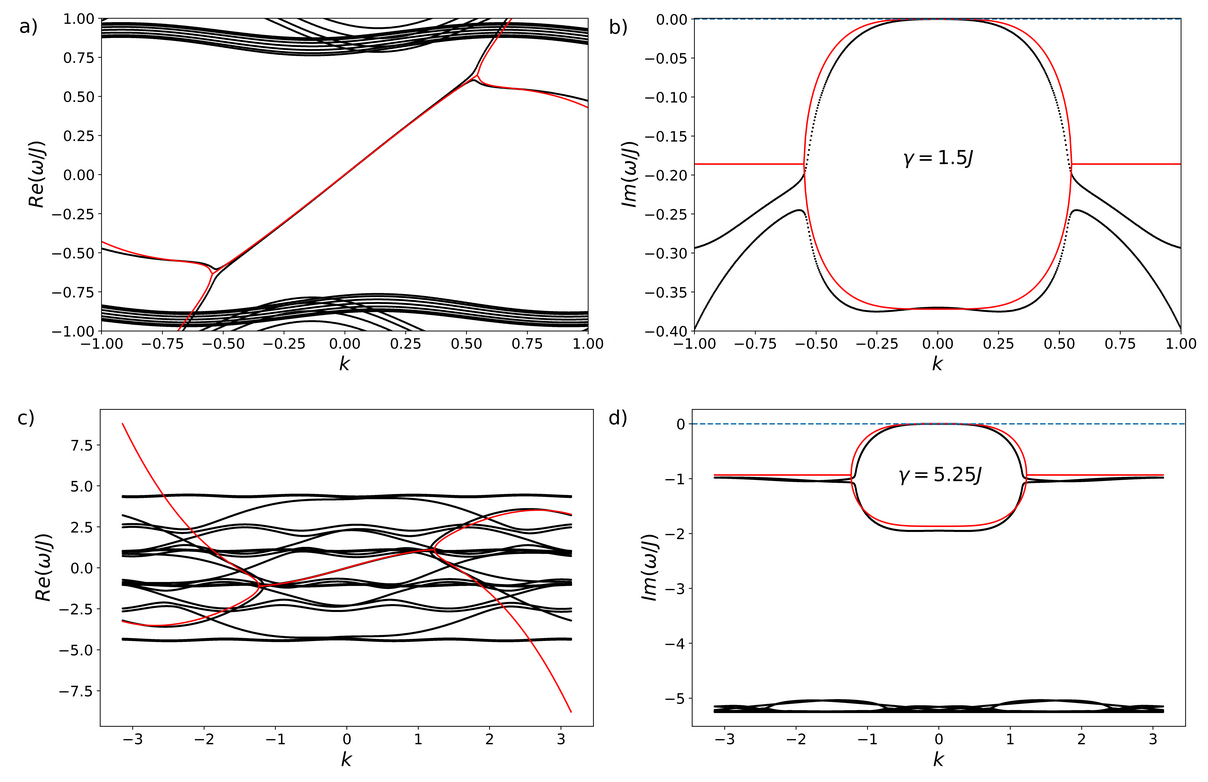}}
    \caption{Real (a),(c) and imaginary (b),(d) part of the elementary excitations spectrum for the 2D model (black dots) and the gGPE model (red lines) for (a),(b) $\gamma= 1.5J$, (c),(d) $\gamma= 5.25J$.
    }
 
\end{figure}

}
\end{multicols}
\end{changemargin}

\end{appendices}

\end{document}